\documentclass[logo,bsc,singlespacing,parskip]{infthesis}
\usepackage{ugcheck}
\pdfoutput=1

\usepackage{microtype} 
\usepackage[style=authoryear-ibid,backend=biber]{biblatex}
\usepackage{graphicx}
\usepackage{subfigure}
\usepackage{pdfpages}
\usepackage{mathtools}

\usepackage{float}

\usepackage{booktabs} 
\usepackage{multirow} 
\usepackage[allcolors=blue]{hyperref} 
\usepackage{wrapfig}
\usepackage{soul}

\begin{document}
\begin{preliminary}

\title{Smart Application for Fall Detection Using Wearable ECG \& Accelerometer Sensors}

\author{Harry Wixley}

\course{Artificial Intelligence and Computer Science}

\project{4th Year Project Report}        

\date{\today}

\abstract{


Timely and reliable detection of falls is a large and rapidly growing field of research due to the medical and financial demand of caring for a constantly growing elderly population. Within the past 2 decades, the availability of high-quality hardware (high-quality sensors and AI microchips) and software (machine learning algorithms) technologies has served as a catalyst for this research by giving developers the capabilities to develop such systems.
This study developed multiple application components in order to investigate the development challenges and choices for fall detection systems, and provide materials for future research. The smart application developed using this methodology was validated by the results from fall detection modelling experiments and model mobile deployment.
The best performing model overall was the ResNet152 on a standardised, and shuffled dataset with a 2s window size which achieved 92.8\% AUC, 87.28\% sensitivity, and 98.33\% specificity. Given these results it is evident that accelerometer and ECG sensors are beneficial for fall detection, and allow for the discrimination between falls and other activities.
This study leaves a significant amount of room for improvement due to weaknesses identified in the resultant dataset. These improvements include using a labelling protocol for the critical phase of a fall, increasing the number of dataset samples, improving the test subject representation, and experimenting with frequency domain preprocessing.

}

\maketitle

\newenvironment{ethics}
   {\begin{frontenv}{Research Ethics Approval}{\LARGE}}
   {\end{frontenv}\newpage}

\begin{ethics}
This project obtained approval from the Informatics Research Ethics committee.\\
Ethics application number: 18048\\
Date when approval was obtained: 2022-01-17\\
%
This project required human participants to simulate falling over while wearing the relevant hardware.
The participant's \hyperref[chap:ethics-info]{\emph{information sheet}} and \hyperref[chap:ethics-consent]{\emph{consent form}} are included in the appendix.\\

\standarddeclaration
\end{ethics}

\begin{acknowledgements}
I would like to thank Kianoush Nazarpour for supervising this project and providing me with invaluable guidance throughout this year, and my fellow students/volunteers who helped me collect data for my custom dataset.

Additional thanks go to my family, friends, and loved ones, for their unconditional
support.
\end{acknowledgements}

\tableofcontents
\end{preliminary}


\chapter{Introduction}

This project has developed a smart mobile application that connects to accelerometer and ECG sensors and uses this live sensor data for fall detection. The application was optimised by performing experiments on the utility of different sensors for fall detection, the quality of varying preprocessing techniques for working with this type of data, and the performance of varying machine learning models for fall detection and prevention.

\section{Contributions}

\subsection{iOS Data Collection Application}
A data collection system is not only important for effective app development (collecting data from the exact same system it will be used on) but also to allow for easy tweaks and data manipulations (ie. adding features) throughout the development cycle.

An iOS application was developed to record data from iPhone movement sensors (accelerometer, magnetometer, gyroscope), and Polar H10 sensors (accelerometer, ECG). This app transmits the data locally using custom requests in Swift to a NodeJS server. This server uses IPv4 based white-listing for security, and connects to a MongoDB instance for data storage.

\subsection{Fall Detection Dataset}
Using the data collection application a dataset for fall detection was collected. The data was collected from 5 subjects with an average of 13.4 recordings per subject, an average recording duration of 117s, and an average of 61 falls per subject.

\subsection{Evaluation of Preprocessing Techniques for Fall Detection Data}
The quality of varying data shift and scaling techniques (normalisation, standardisation, and log transform) was evaluated on this dataset. Overall, standardisation proved to be most effective.

The usage of other preprocessing techniques was determined by inspecting the nature of this data (time series sensor data), and reviewing existing literature that work with similar datasets and/or use-cases. Given this data is a time-series sliding windows were used for formatting the data into samples, and labelling lag was utilised to adjust the training labels for fall prevention models.

\subsection{Evaluation of ML Models for Fall Detection \& Prevention}
The performance of varying machine learning models on this dataset was evaluated for fall detection and prevention.

\subsubsection{Fall Detection}
Baseline experiments\footnote{These included experiments on the following models: Linear Regression, Decision Trees, Support Vector Machines, K-Nearest Neigbours, Single-Layer Neural Networks, Bernoulli Naive Bayes, and Gaussian Naive Bayes.} were performed and the optimal model found was KNN which achieved a test AUC of 72.17\%. Experiments for varying deep learning architectures (CNNs and LSTMs) were also performed and the optimal model found was the ResNet152 which achieved a test AUC of 92.80\%.

\subsubsection{Fall Prevention}
While the focus on this project was for fall detection an attempt was made to test whether data could be used for prevention. This was evaluated on the best performing detection model (ResNet152 on data with 2s window sizes) for 100ms and 200ms labelling lag. However, in all cases only very small progressions in validation AUC were made (the maximum validation AUC achieved was 51.58\%) and thus these results were not included.

\subsection{iOS Fall Detection System}
A prototype commercial iOS fall detection app was developed to test the fall detection models and provide a platform for model testing along with other commercial features.

The application was configured with a Firebase backend (allowing user login and create account functionality), user fall notifications (to notify the user if a fall is detected), and automatic emergency contact notifications (to notify a user's emergency contacts if a fall is detected and the user is unresponsive).

\section{Technical Novelty}

Unlike the majority of existing fall detection models, user-specific features including height, and weight were integrated into the machine learning model input. This use of personalised data is very valuable as it allows the model to become more fine-tuned on a per user basis making it more reliable when new users start using the product. It also allows the model to be further strengthened when additional personalised data is added to the sample inputs.

\section{Clinical Significance}
In the UK alone, millions of older people (65+) are worried about falling over, with 4.3 million (36\% of the UK's elderly population) putting it at the top of their list of concerns (\cite{ageuk}). It is evident that this is a significant issue given nearly 100,000 elderly people suffered hip fractures in 2017/2018, the majority of those due to falls (\cite{ageuk}). Such numbers do not come at a small cost for NHS Health and Social Care, with an estimated £1 billion annual expenditure on hip fractures alone (\cite{ageuk}).

It is acknowledged that several commercial fall detection systems already exist and thus such a system is not novel. However, an application that can improve the reliability and accuracy of fall detection could make a material difference given the sheer size of the problem.


\section{Rigour}
The data collection application was developed by using robust protocols for data transmission, and by performing a lot of software testing (to ensure all edge cases were handled ie. a data POST request failure).


The iOS fall detection application was developed by using a fast and secure cloud service for the backend, running background processes for live model inference, and by performing a lot of software testing (to ensure all edge cases were handled ie. the background handling of fall notifications).

\chapter{Background}

\section{Relevance}\label{sec:relevance}
Over the last 30 years, developed countries have experienced the increasing challenge of an ageing society as the population of elderly people (aged 65+) has steadily risen (\cite{hsieh_chiang_huang_chan_hsu_2011}). The United Nations even predicts that by 2050 the world's elderly population will have doubled (\cite{un_2019}). This is a problem as during the ageing process people become far more susceptible to psychological, nervous system, or physical defects/diseases, which can directly impede their mobility and put them at higher risk for falling over. Even the fear of falling has been shown to be associated with negative consequences such as a decline in physical and mental performance, an increased risk of falling and a progressive loss in health-related quality of life (\cite{scheffer_2008}).

It is evident that falls are a problem given that they are the leading cause of injury-related death, and the third leading cause of poor health amongst elderly people (\cite{scheffer_2008}). In the UK alone, millions of older people are worried about falling over, with 4.3 million (36\% of the UK's elderly population) putting it at the top of their list of concerns (\cite{ageuk}). 

However, falls are not exclusively problems of the elderly, they can also affect younger people with conditions, diseases and/or disabilities. Seizures, anemia, pregnancy, sport, strokes, heart attacks, and many more conditions can lead to unexpected falls. Each year an estimated 684,000 individuals die from falls globally (of which over 80\% are from low and middle-income countries), and there are 37.3 million other falls that are severe enough to require medical attention (\cite{whofalls}).

Falls do not only pose substantial health issues but they also pose substantial financial issues. The average health system cost per fall injury for elderly people in the Republic of Finland and Australia are US\$ 3611 and US\$ 1049 respectively. In the UK nearly 100,000 elderly people suffered hip fractures in 2017/2018, the majority of those due to falls (\cite{ageuk}). Such numbers do not come at a small cost for NHS Health and Social Care, with an estimated £1 billion annual expenditure on hip fractures alone (\cite{ageuk}). The medical cost of fall-related related injuries in the US in 2020 was projected to reach around US\$ 32.4 billion (a US\$ 12.2 billion expenditure increase from 1994) (\cite{rajagopalan_litvan_jung_2017}).

It is evident that falls are a serious problem globally not only for people's health but also their finances. Thus developing a cost-effective system that could provide timely support to user's who have fallen over could pose massive benefits.

Developing a fall model for prevention rather than detection would clearly be preferable given this would allow us to prevent user injury altogether. However, such a task is extremely difficult due to the fact that fall events are fast and stochastic, and thus such a prevention model has never been successfully developed.

\subsection{Google Trends for Fall-related Queries Over Time}\label{subsec:google_trends}
\begin{figure}[H]
    \centering
    \includegraphics[width=0.6\textwidth]{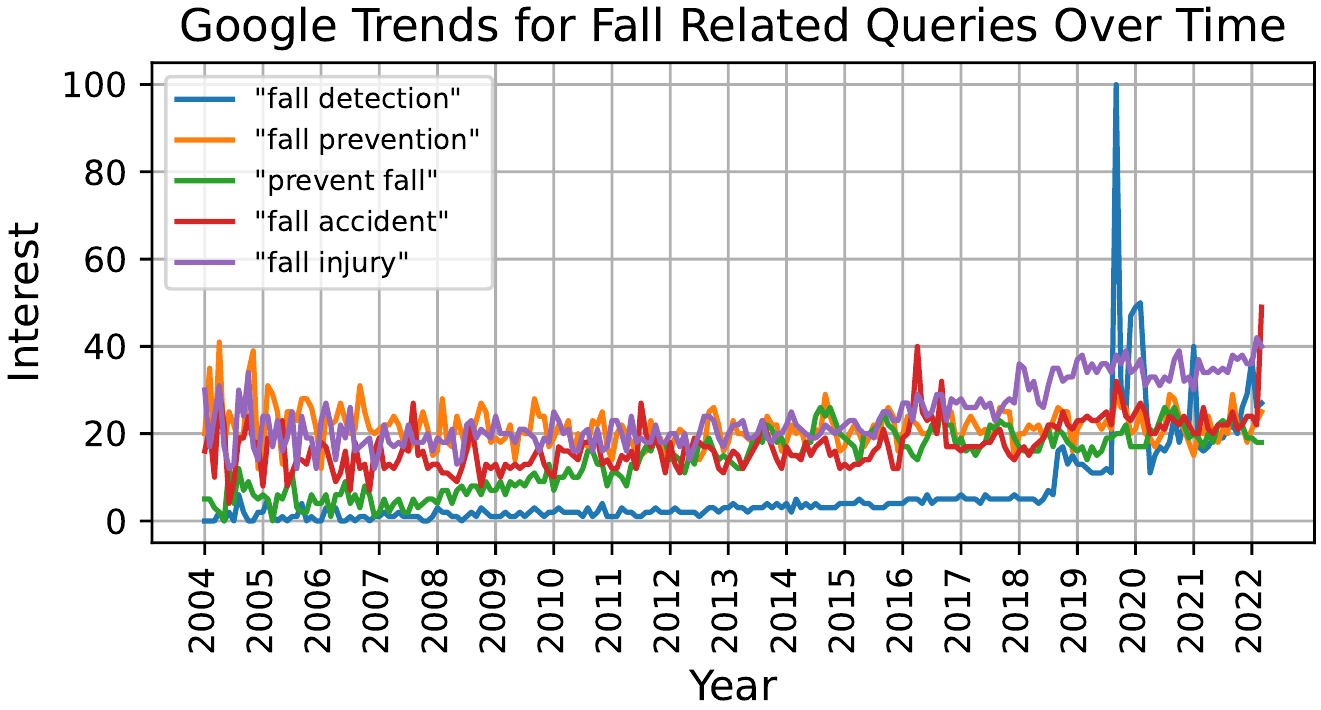}
    \caption{Interest of fall related web search queries from January 2004 to March 2022. This data is taken from Google Trends by comparing the relevant queries (shown in the plot's legend) across all categories and regions. Interest is calculated by dividing each data point by the total searches of it's geography and time range it represents to compare relative popularity, otherwise, places with the most search volume would always be ranked highest. The values are then normalized with the maximum interest, such that the highest interest is 100.}
    \label{fig:search-trends}
\end{figure}
From the chart above we can see that there has been consistent interest in fall prevention for a very long time, unlike fall detection. It is possible this was due to the fact users wanted something to prevent falls, however, they were unaware of the existence/capabilities of a fall detection system (which could be due to the fact no commercial fall detection systems existed yet). This is supported by the fact that in the past 4 years we can see there has been a significant interest increase in fall detection systems, which is potentially explained by Apple's commercialization of such a technology on their smart watch in September 2018. On this watch, accelerometer and gyroscope sensors were employed for hard-fall detection (\cite{apple_watch_2018}).

\subsection{Industry Relevance}
The use of IoT devices in the medical field is becoming more and more relevant:
“The rapid growth of the global market for the Internet of Things of Medical Devices (IoMT), improved confidentiality, high speed of real time data processing, significant reduction of technology cost, and improved quality of outpatient medical services drive a global paradigm shift in healthcare.” - \cite{izotov_2021}

Given this rapid growth of medical IoT devices and the relevance of fall detection (as discussed in sections \ref{sec:relevance} and \ref{subsec:google_trends}) it is evident this type of system has a significant demand in industry.


\section{Usability}
When thinking about developing such a system we must consider the implications this has for a given user, and how their behaviour may affect its efficacy.

One of the most significant problems with being able to detect falls is that it typically requires users to wear large and power-hungry devices, this is obviously not ideal as it requires a user to wear it all the time, and regularly charge it. This leaves a large responsibility on the user (who will be mainly be elderly) who may forget to wear it for the day, charge it, or even turn it on. 

Furthermore, given this will be used for a safety-critical system we must be particularly wary of the hardware issues our device's sensors can be susceptible to. For example, gyroscope sensors are susceptible to becoming poorly calibrated if exposed to fluctuations in temperature and/or humidity (\cite{shu_2021}). Such calibration issues could be extremely detrimental to the robustness of the system as they would result in the transmission of inaccurate signals and thus inaccurate fall detection predictions.

\subsection{External Detector}
A possible solution to the size and power problem would be to create an external fall detection system that can detect if a person falls over in a fixed space (such as a home for elderly people). This would make it easier to configure more power-hungry devices as they could remain plugged in constantly and would not need to be fitted into a small wearable device. Being able to use more power-hungry devices would allow us to use more complex machine learning models for our predictions to improve our detection accuracy, and also compute model predictions faster.

Configuring an external system for a given room could be achieved using multiple LiDAR\footnote{Light Detection and Ranging} sensors or cameras located in each of the corners of the ceiling which would be used to localise a person's position in the room.

\subsection{Wearable Detector}
Despite the various issues of wearable detectors they still have a valid use-case as they allow us to monitor a user constantly throughout the day no matter where they go. This is unlike external sensors which are limited to indoor spaces that are already setup with this system. Not only would this be difficult to setup on a large scale (ie. a house with many irregular shaped rooms) but it can also become very costly as each room would need it's own set of LiDAR/camera sensors.

\subsubsection{Smart Watch}
Smart watches are useful as many have integrated biometric (ie. ECG) and movement (ie. accelerometer, gyroscope) sensors. The new Samsung smart watch is particularly exciting due to it's collection of blood pressure data which could prove valuable for fall detection (\cite{samsung_2022}).

However, due to the complexity of fitting falls using an accelerometer on the wrist (given arms are obviously involved a very wide array of actions that are independent from falling) I decided against using this hardware.

\subsubsection{Insoles/Socks with Pressure Sensors}
This method would be extremely useful for both abnormal gait\footnote{Gait is a person's pattern of walking.} detection (can easily estimate cadence, walking asymmetry and other useful statistics) and fall detection (can tell if a person falls over when there is no pressure on the insoles). These pressure sensors could even be used in conjunction with accelerometer and gyroscope sensors on a phone (to improve the accuracy as much as possible) which would also allow all the processing to be done on the phone meaning no extra processing device would be needed.

After much deliberation and investigation this method was not used due to the hardware acquisition costs (which could also be an obstacle to a commercial application). After extensive research the cheapest sock/insole hardware found with an iOS SDK available was \href{https://store.sensoriafitness.com/sensoria-core-pair/}{the Sensoria smart sock} which costs £400.

\subsubsection{Mobile App}
Another solution would be configuring a fall detection system in an app on a mobile device. Since the majority of people already have mobile phones which are always turned on and carried everywhere it would require no extra moving parts. Furthermore, a lot of new phones already have an array of integrated high-quality movement sensors that can be interfaced with taking away a lot of the hardware limitations we may face.

However, the major limitation of using sensors from a mobile phone for fall detection is that the phone's location is never fixed (ie. it could be held in someone's hands, put in a bag, dropped on the floor etc.) which would be extremely unreliable for fall detection. Given these limitations phone sensors were not used for fall detection modelling.

\subsubsection{Electrical Heart Sensors}
Electric heart sensors allow us to collect electrocardiogram\footnote{Electrocardiogram is one of the simplest and fastest tests used to evaluate the heart (\cite{electrocardiogram_2022}).} (ECG) data. This data would be potentially valuable for both abnormal gait detection and fall detection. It would help abnormal gait detection as it has been found that for Parkinson's patients during a freezing of gait\footnote{Freezing of gait (FOG) is a disabling, episodic gait disturbance that is common among patients with advanced Parkinson's disease (PD). FOG typically lasts a few seconds during which time the patient feels as if his or her feet are glued to the ground.} there is an observable change in their heart rate which could be used for detecting falls (\cite{maidan_plotnik_mirelman_weiss_giladi_hausdorff_2010}). We can expect heart rate to play a significant role in fall detection as a person may get a fright or be in shock after a fall resulting in a raised heart rate. It is also possible ECG could be used for fall prevention (\cite{melillo__2015}).

This method was used employing a Polar H10 device which incorporates both accelerometer and ECG sensors, and has an iOS SDK enabling interface with these sensors via an iOS app. Further details on the reasoning for this is discussed in chapter \ref{chap:data-coll}.

\section{Domain Specific Research}

\subsection{Phases of a Normal Gait}
Given gait abnormalities can directly increase a person's chance of falling over it is important to identify what a normal gait looks like:

\begin{figure}[h]
    \centering
    \includegraphics[width=0.8\textwidth]{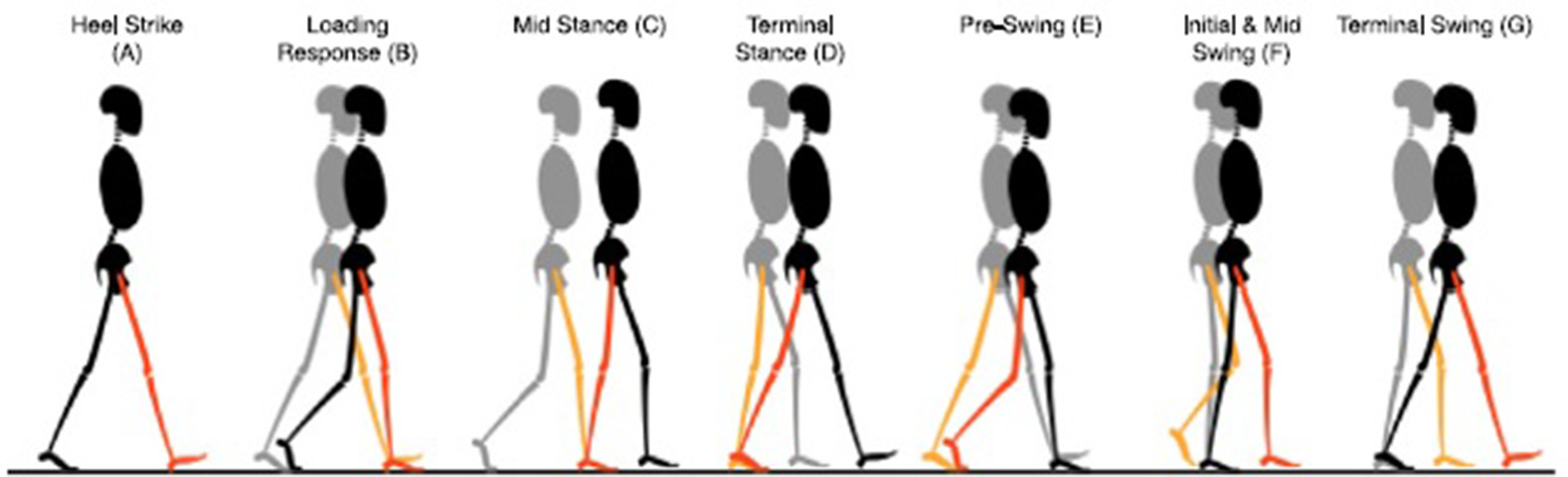}
    \caption{Phases of a normal gait (\cite{shrivastava_bharti_pateriya_2021})}
    \label{fig:my_label}
\end{figure}

\subsection{Anatomy of a Fall}
A fall happens when the center of gravity (CG) of an individual's trunk becomes skewed with the foundation of their feet on the floor. A person's CG is typically at the level of their sternum foremost to the spine, at which all the weight of their torso is evenly distributed.

\begin{figure}[h]
    \centering
    \includegraphics[width=0.5\textwidth]{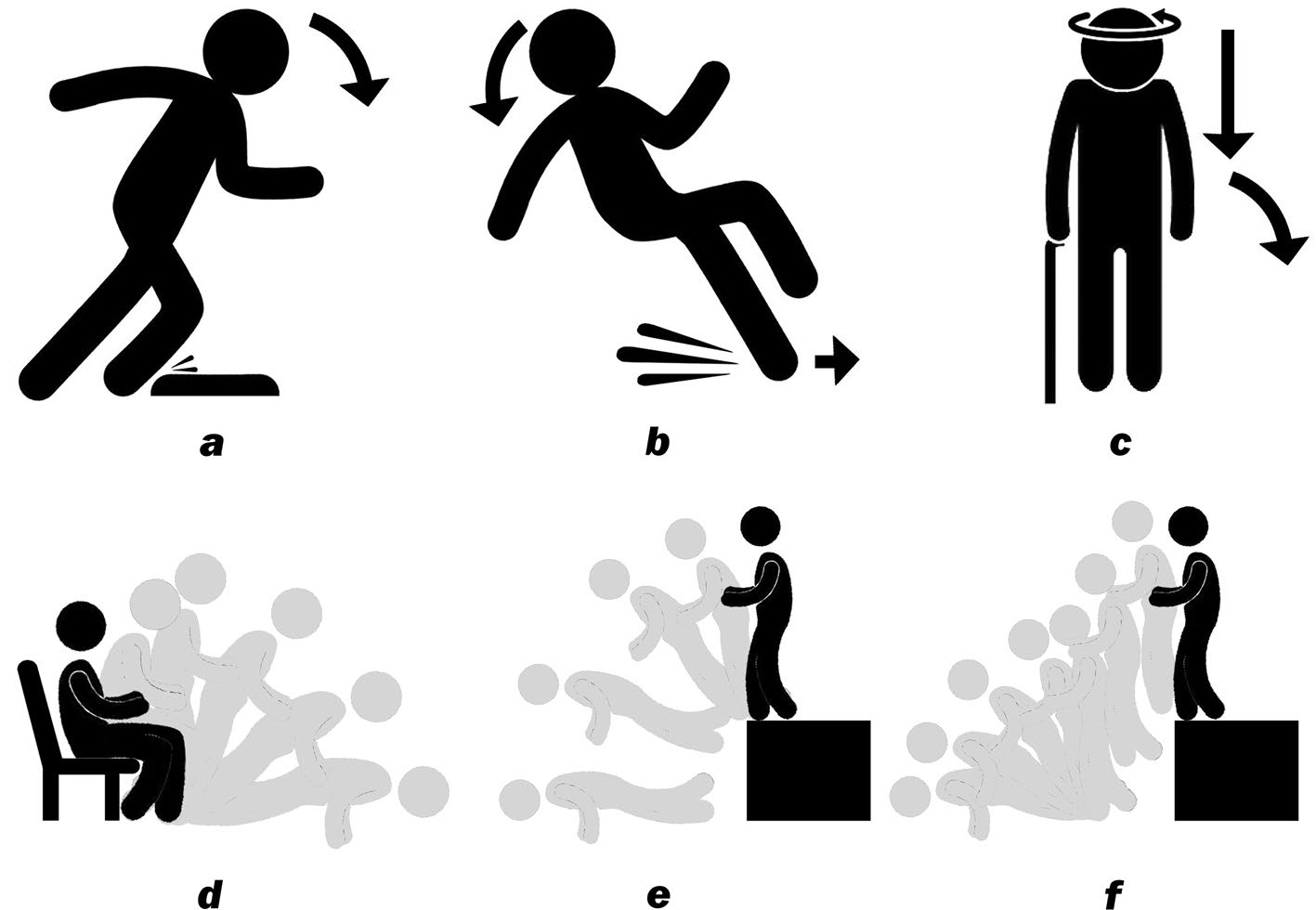}
    \caption{Selected fall types. (a) Stumbling. (b) Slipping. (c) Fainting. (d) Getting up from a sitting position (i.e., a chair) and falling. (e) Falling from a high structure (i.e., stairs, ladders, etc.). (f) Jumping down from a high structure and falling. (\cite{shu_2021})}
    \label{fig:anatomy_of_a_fall}
\end{figure}

\subsubsection{Types of Falls}

\noindent
\begin{itemize}
    \item Stumbling (Fig. 2.2a) happens when a person comes into contact with an unperceived object. This point of contact is slowed down by the object, however, the person's inertia keeps their CG moving thus resulting in them falling over due to a misalignment between their CG and foundation. Stumbling typically occurs in poorly lit rooms with misplaced items on the floor. People with neurological or musculoskeletal issues are more in danger of stumbling.
    \item Slipping (Fig. 2,2b) happens when the frictional force between a person's foot and the floor is overcome by the force of the person's foot connecting with the floor. This inertia results in the person's feet to start moving away from their CG resulting in a misalignment between their CG and foundation causing them to fall. This type of fall is particularly prevalent in elderly people because of the reduced density of sensorimotor nerve strands in their feet, and individuals with preexisting gait issues due to their asymmetrical walking patterns. Inappropriate footwear and environmental factors (wet walkways from rain, frozen walkways from snow, etc.) can also further improve the probability of this type of fall.
    \item Fainting (Fig. 2.2c) is a result of weakened cerebral perfusion and transient brain hypoxia, causing a deficiency of postural tone. It is characterized by a direct plummet of the head and torso, while the person's CG stays in accordance with their feet, this is then followed by a bending of the torso and knee(s), this imbalance results in the whole body stumbling and collapsing. Any pathology hindering enough oxygenated bloodstream to the mind can bring about fainting, this could range from chronic anemia, vasovagal syncope, paroxysmal arrhythmia, to dysautonomia, just to give some examples.
    \item Other typical types of falls are variations of stumbling and slipping. For example, falling after attempting to sit in or get up from a chair (Fig. 2.2d) is commonly observed when the elderly use lightweight chairs or stools with wheels. Another example includes falling down from a high structure such as a ladder, desk, or a set of stairs (Fig. 2.2e). And lastly, falling after attempting to jump down from a high structure (Fig 2.2f).
\end{itemize}

\subsubsection{Fall Phases}

\begin{figure}[h]
    \centering
    \includegraphics[width=0.8\textwidth]{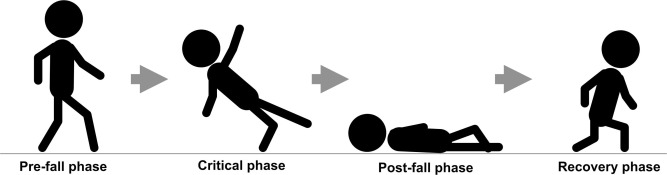}
    \caption{Phases of a fall (\cite{yajai_rasmequan_2017})}
    \label{fig:phases_of_a_fall}
\end{figure}

Given the various phases associated with a fall it is very important in how we choose to label our data as to represent this fall phase.

For example, we must choose whether it would be worthwhile to label all these phases as a fall. However, given falls occur over varying durations it is possible only certain phases will be present in a given sample (if the window size is smaller than the fall duration), and thus it is important to identify if this could be problematic for classification (if some fall phases are similar to any ADLs \footnote{Activities of daily living}).

One problem for including all phases as a fall, means these individual phases all get classified as a fall. This is problematic as the post-fall phase could prove to be very similar to lying down on a bed/couch which is evidently not a fall. Given the fact that the critical phase must occur in order for the post-fall phase to occur it would be most useful/robust to just identify the critical phase of a fall. The difficulty with doing this though is creating a labelling technique which only identifies this critical phase. Otherwise, another solution to this problem would be to label the critical phase and post-fall phases as falls but use larger window sizes so the model can see the combination of these phases in order to identify a fall. This is likely to produce a more robust/accurate model given the increased amount of sample input data. The downfalls of such a technique are obviously the complexity given we have to increase the window size rather dramatically (to include an entire fall) and thus the size of inputs to the model.

\section{Literature Review}
There are various review papers that give an account of the development of fall detection from different aspects. Due to the rapid development of smart sensors and related analytical approaches, it is necessary to illustrate the trends and development frequently. In order to gauge the progression of fall detection research I will review the key pioneering papers, and the overall top 3 cited papers.

\begin{figure}[H]
    \centering
    \includegraphics[width=0.6\textwidth]{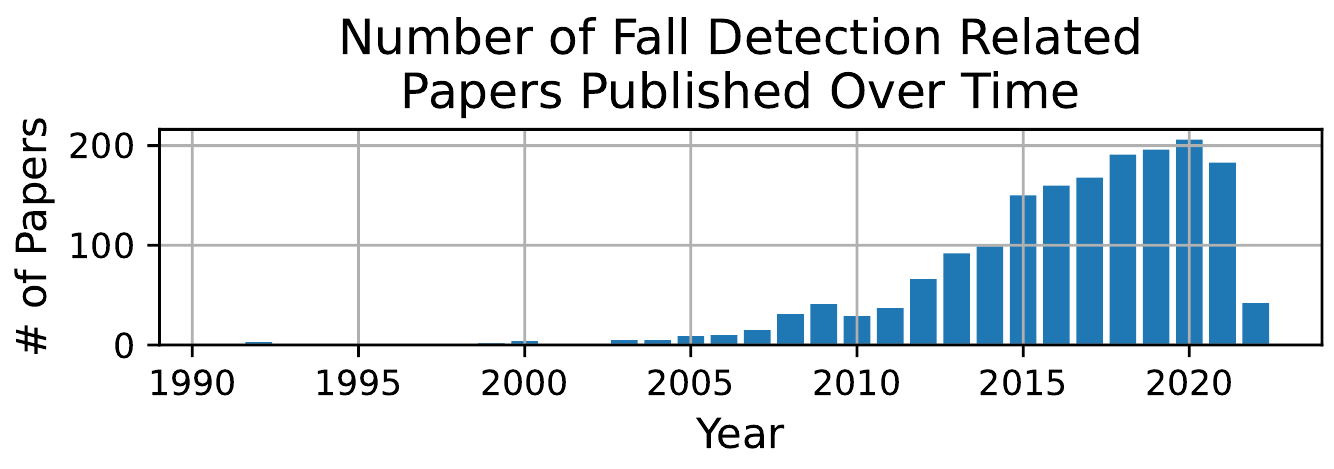}
    \caption{Number of fall detection related papers published over all time (1991-2022). This data was collected from Web of Science by querying all publications that satisfied the following relation: 
AK=(("fall" OR "falls" OR "falling") AND ("detection" OR "detector" OR  "detect" OR  "detecting" OR "monitor" OR  "monitoring")) (where AK refers to the author's keywords for their given paper).}
    \label{fig:pub-trends}
\end{figure}

\subsection{Key Results of Pioneering Papers}
\textbf{\citetitle{lord_colvin_1991} by \cite{lord_colvin_1991} [29 Citations]:}\\
The aim of this paper was to attempt the detection and quantification of subject falls within normal living areas. This was done by performing 2 experiments with differing fall data acquisition techniques: using video monitoring, and using a wearable badge with an embedded accelerometer and analog-input microcomputer chip.

For the video experiment it was concluded that there was a practical means of detecting and evaluating falls through video data without continuous monitoring of the video monitors by personnel. However, this system was rejected by a private nursing home and a major hospital (John Hopkins) due to privacy concerns.

For the wearable experiment it was concluded that the badge would be a more viable means for fall detection due to the lack of privacy concerns, and the viability of using this badge/accelerometer for detecting falls. The badge was capable of measuring forces over a preset limit, and logging the time and duration of accelerations. This data could then be loaded onto a desktop for statistical analysis and threshold optimisation for fall detection.

\textbf{\citetitle{williams_doughty_cameron_bradley_1998} by \cite{williams_doughty_cameron_bradley_1998} [135 Citations]:}\\
In this paper the design of a smart sensor to detect falls and monitor activity is discussed in terms of its integration within an intelligent telecare system. The proposed device consisted of a piezoelectric shock sensor to detect the impact, a mercury tilt switch to monitor the orientation of the client, and a microchip for data processing.

It also suggested the possibility of developing fall detection algorithms with integrated user-specific parameters such as age, gender, mobility, medication, cognitive function, activities of daily living, and history of previous falls.




\subsection{Key Results of Top 3 Cited Papers}

\textbf{\citetitle{yang_hsu_2010} by \cite{yang_hsu_2010} [566 Citations]:}\\
This paper reviews the quality of accelerometry-based wearable devices for physical activity monitoring user-cases such as fall detection.

It was identified that balance control or postural stability of the body while standing still or walking has been regarded as an important predictor of risk of falling of the elderly, and that it could be measured by using accelerometers placed at the back of a subject.

\textbf{\citetitle{bourke_2006} by \cite{bourke_2006} [494 Citations]:}\\
In this paper simulated falls and activities of daily living (ADL) were performed by elderly subjects in order to investigate the ability to discriminate between falls and ADLs using accelerometer sensors on the trunk and thigh. Fall detection algorithms for these sensors were devised using threshold techniques. 

Overall, the accelerometer located on the trunk achieved the best detection performance with 100\% specificity on the upper fall threshold and 91.25\% specificity on the lower fall threshold.

\textbf{\citetitle{mathie_2004} by \cite{mathie_2004} [491 Citations]:}\\
This paper reviews the use of accelerometer-based systems in various movement monitoring use-cases: gait, sit-to-stand transfers, postural sway, and falls. The scope and applicability of such systems in unsupervised monitoring of humans are considered. The authors even suggest the possibility of integrating these different system use-cases to provide a more comprehensive system that is suitable for measuring a range of different parameters in an unsupervised monitoring context with free-living subjects.

Accelerometer sensors were particularly favoured by these authors due to their suitability for long-term monitoring of free-living subjects because it can provide objective, reliable monitoring of unconstrained subjects for low cost.

\subsection{State-of-the-art Solutions}
The most effective solution for vision-based fall detection to date was developed by \cite{sehairi_2018} using an ANN\footnote{Artificial Neural Network} on RGB camera inputs. This system was able to predict
the correct class with an accuracy that can reach up to 99.61\% with a maximum global error of 1.5\%.
However, this solution has the problem that it is bounded to an internal space, can be very expensive to implement (requires a camera in every room), and has privacy concerns.

The most effective solution for wearable-based fall detection to date was developed by \cite{zhang_li_wang_2021} using the ResNet architecture with a customised output threshold moving method (to handle the class imbalance ratio) on the public SisFall dataset (which employs accelerometer and gyroscope sensors). This system was able to make predictions with AUC 98.35\%, sensitivity 99.3\%, and specificity of 91.86\%.

\section{Existing Technologies} 

There are a variety of different commercial fall detection systems available, however, the majority of them are wearable devices due to the privacy concerns of vision-based systems. These wearable systems all typically employ basic movement sensors such as accelerometers, and gyroscopes.\\

\textbf{\underline{Some of the most popular commercial systems (\cite{pickavance_2021}):}}
\begin{itemize}
    \item \underline{Apple Watch:} Apple has already introduced a fall detection system in their CoreMotion library that works on WatchOS.
    \item \underline{SureSafeGO:} a pendant with GPS and a button that once clicked connects you to an operator at the SureSafe response center for assistance. 
    \item \underline{GreatCall Lively:} a watch with a button that once clicked contacts emergency services via your phone.
\end{itemize}

\chapter{Data Collection}\label{chap:data-coll}

\section{Hardware}

\subsection{iPhone}
Requires iOS 15.2 with accelerometer, three-axis gyroscope, and magnetometer sensors.

\subsection{ECG Sensor}
Polar H10 Heart Rate Sensor, firmware version 3.3.1. This device costed £76.50.

This device was chosen mainly due to it's combination of high-quality ECG and accelerometer sensors, and the availability of an iOS SDK (allowing this sensor to be interfaced with via Bluetooth). This device also boasts a battery life of 400 hours, and can operate at temperatures from $-10^{\circ}$C to $+50^{\circ}$C making it very suitable for everyday usage. It boasts extra interference-preventing electrodes and slip-preventing silicone dots making it even more robust for data collection (\cite{polar_sensors}).

\section{Software}

\subsection{Interfacing with Hardware Sensors}

\subsubsection{iPhone}
\href{https://developer.apple.com/documentation/coremotion}{Apple's CoreMotion Swift library} was used for interfacing with the iPhone's various movement sensors (accelerometer, gyroscope, magnetometer). 

This library allowed me to create a MotionManager object which could start and stop different movement services. These services were pushed onto DispatchQueues (which employ process concurrency) allowing executions to be efficient as possible.

\subsubsection{Polar H10 Heart Rate Sensor}
\href{https://github.com/polarofficial/polar-ble-sdk}{Polar's Swift SDK} (version 3.2) was used to interface with the Polar H10 device.

This SDK allowed me to use the \href{https://developer.apple.com/documentation/corebluetooth}{CoreBluetooth} library to connect/disconnect to the H10 device, retrieve device information (such as device ID and battery life), and stream data from the accelerometer and ECG sensors.

\subsection{User Interface}
Storyboard was used in Xcode in order to develop a simple interface that would ensure easy and robust data collection. 

\subsubsection{App pages:}

\begin{figure}[ht]
\centering
\subfigure[Connection]{%
\includegraphics[width=0.15\textwidth]{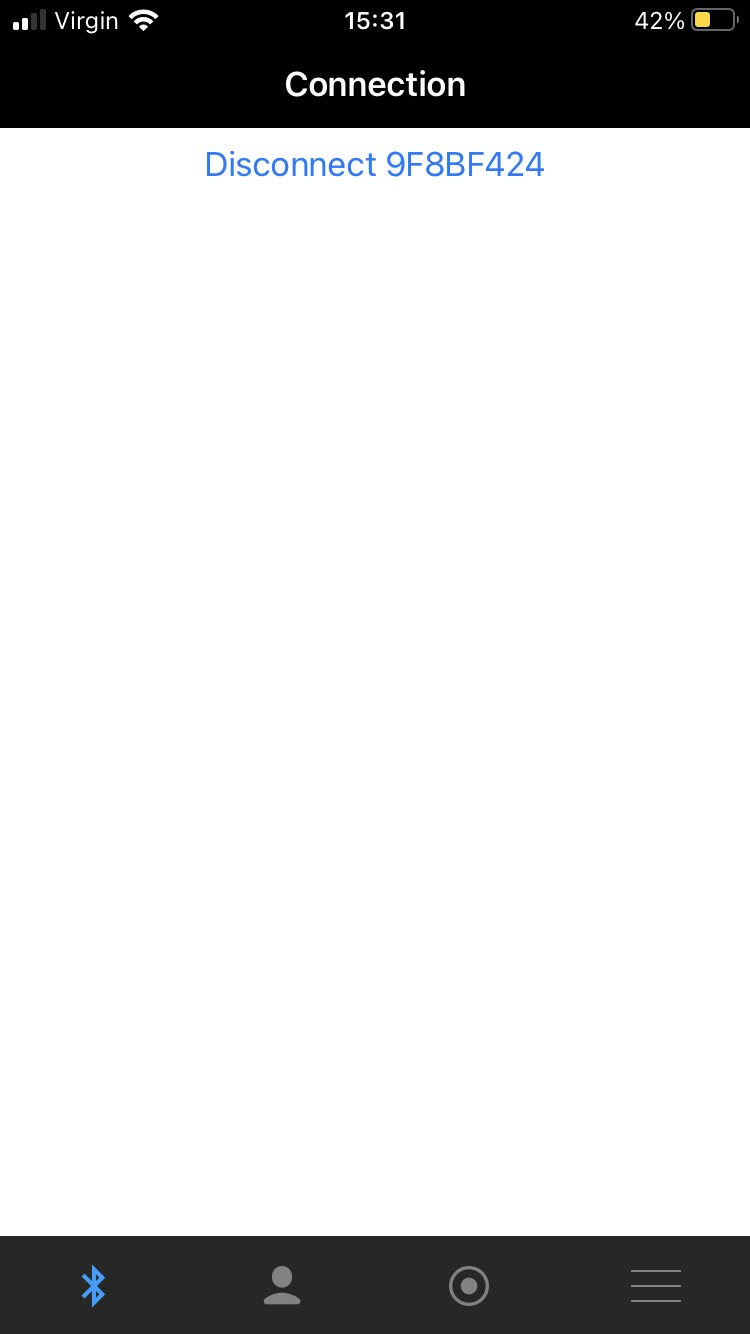}
\label{fig:lhfddg-connection}}
\quad
\subfigure[Add a subject]{%
\includegraphics[width=0.15\textwidth]{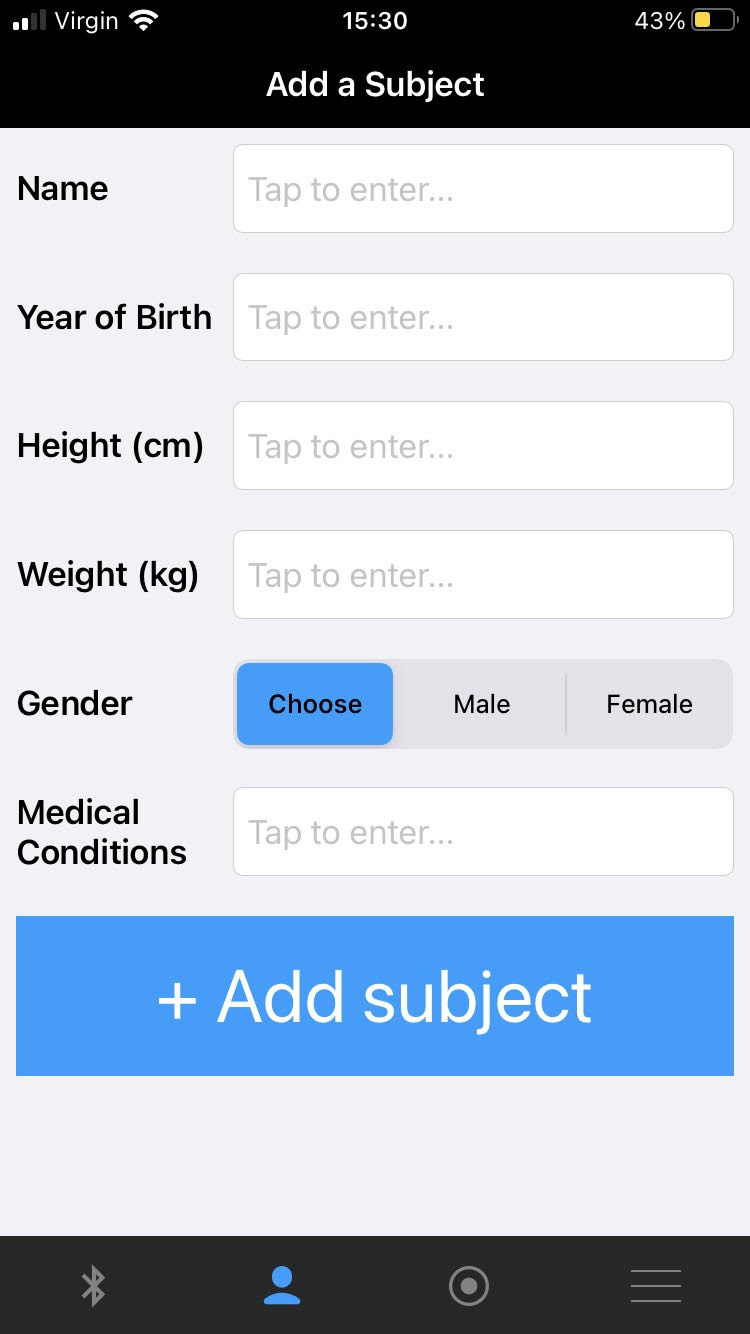}
\label{fig:lhfddg-addsubject}}
\quad
\subfigure[Create a recording]{%
\includegraphics[width=0.15\textwidth]{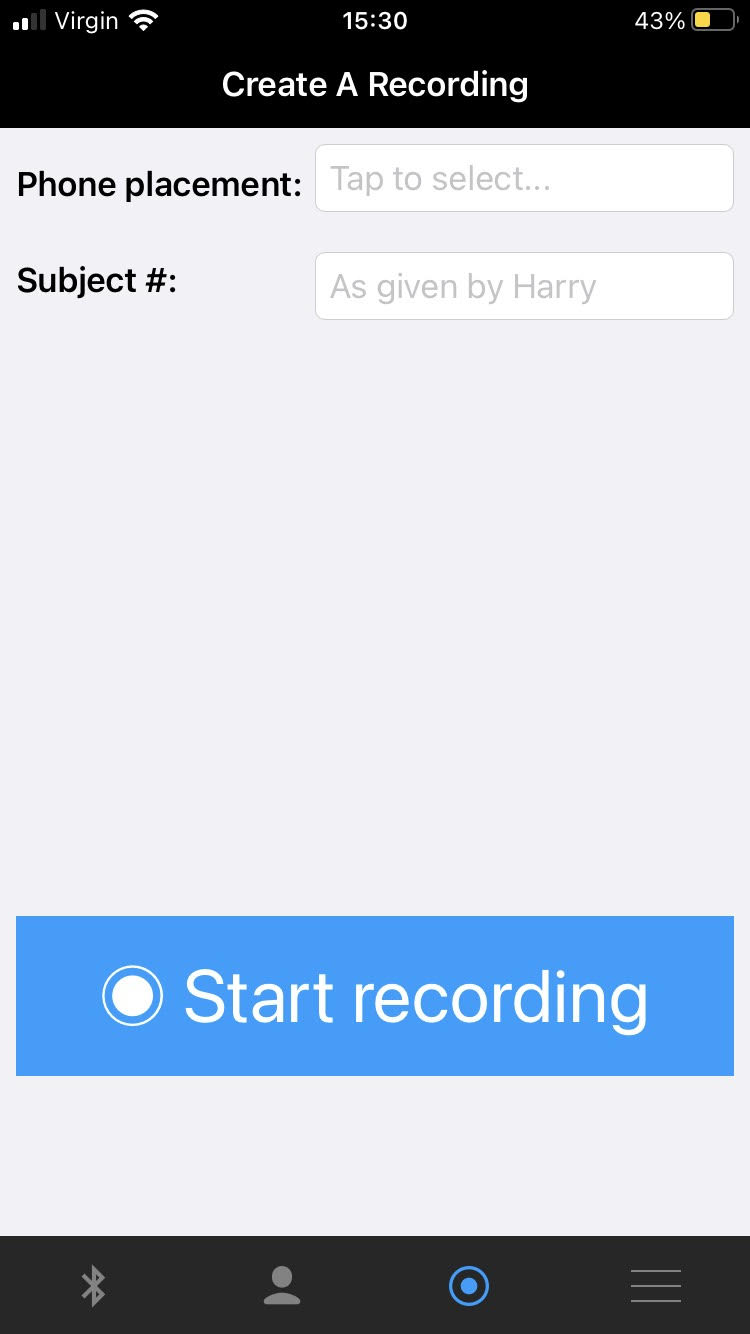}
\label{fig:lhfddg-createrecording}}
\quad
\subfigure[Menu]{%
\includegraphics[width=0.15\textwidth]{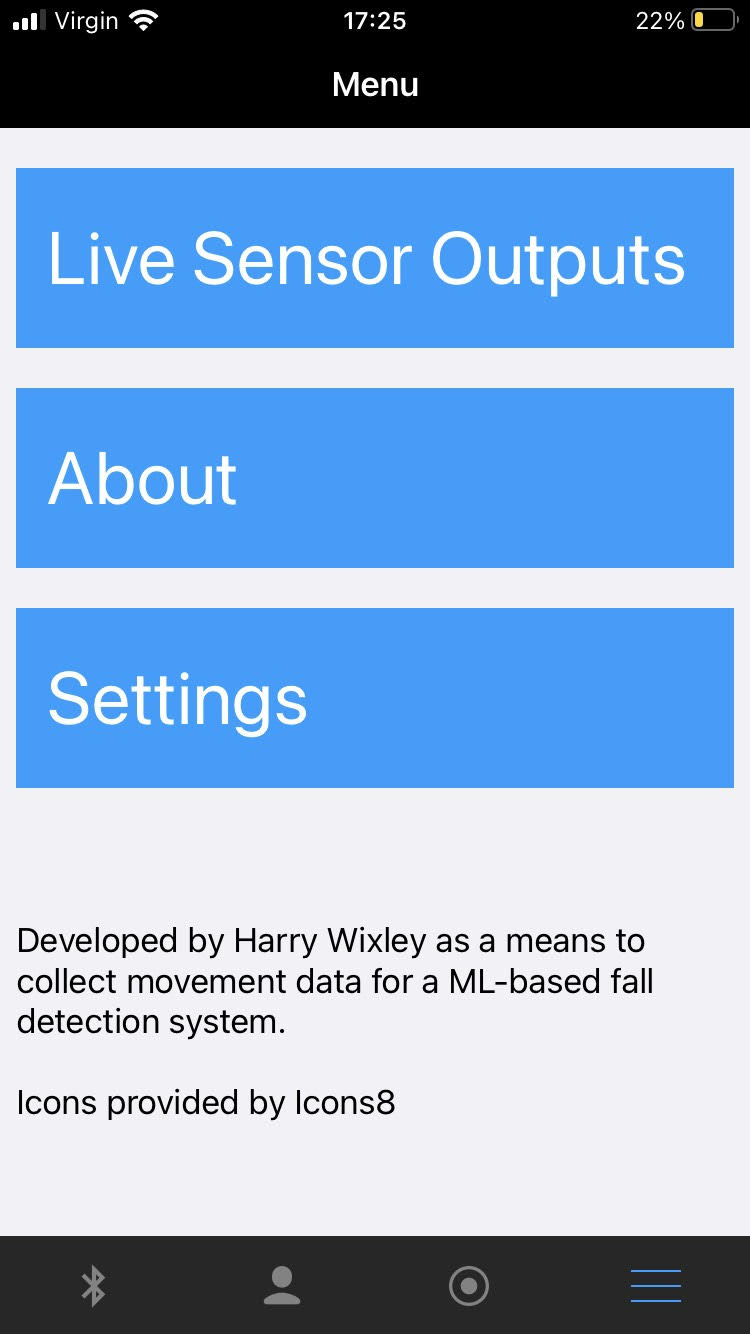}
\label{fig:lhfddg-menu}}
\caption{Main app pages}
\label{fig:lhfddg-main}
\end{figure}

\begin{figure}[ht]
\centering
\subfigure[Recording in progress]{%
\includegraphics[width=0.15\textwidth]{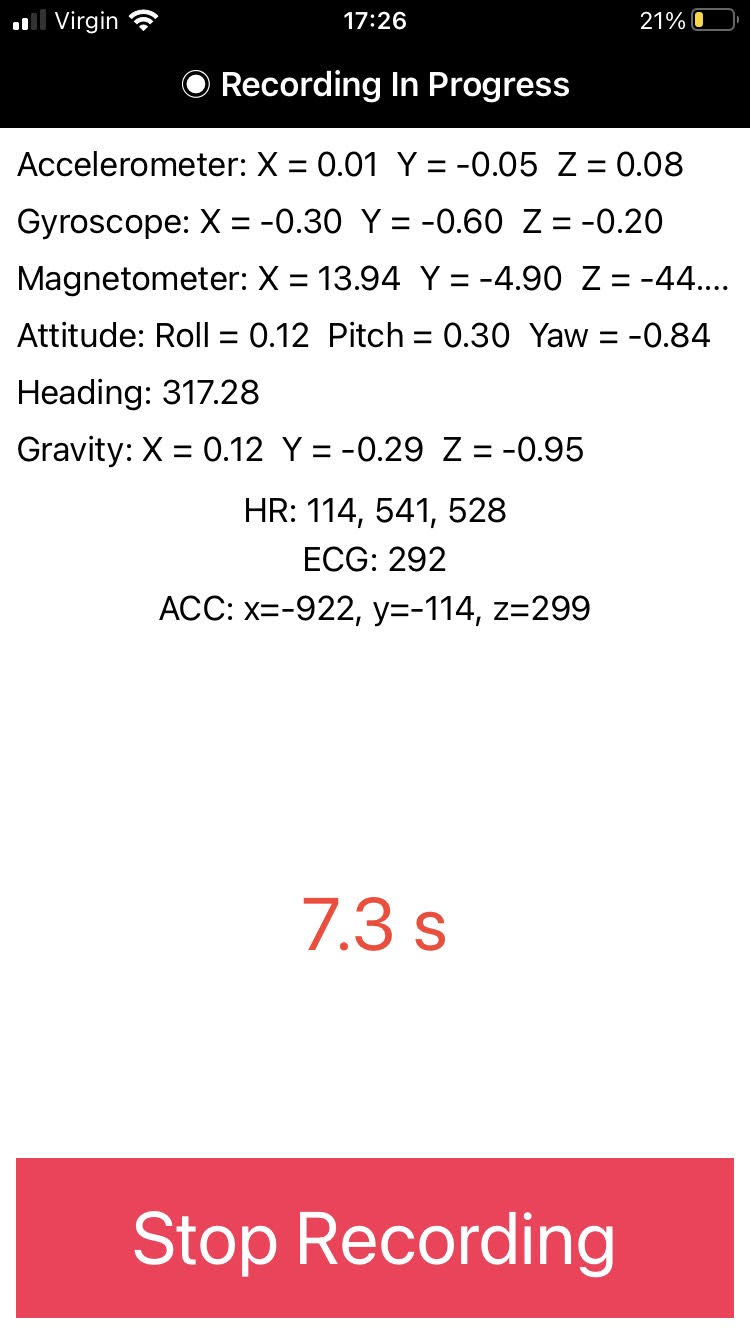}
\label{fig:lhfddg-recordingprogress}}
\quad
\subfigure[Recording stopped]{%
\includegraphics[width=0.15\textwidth]{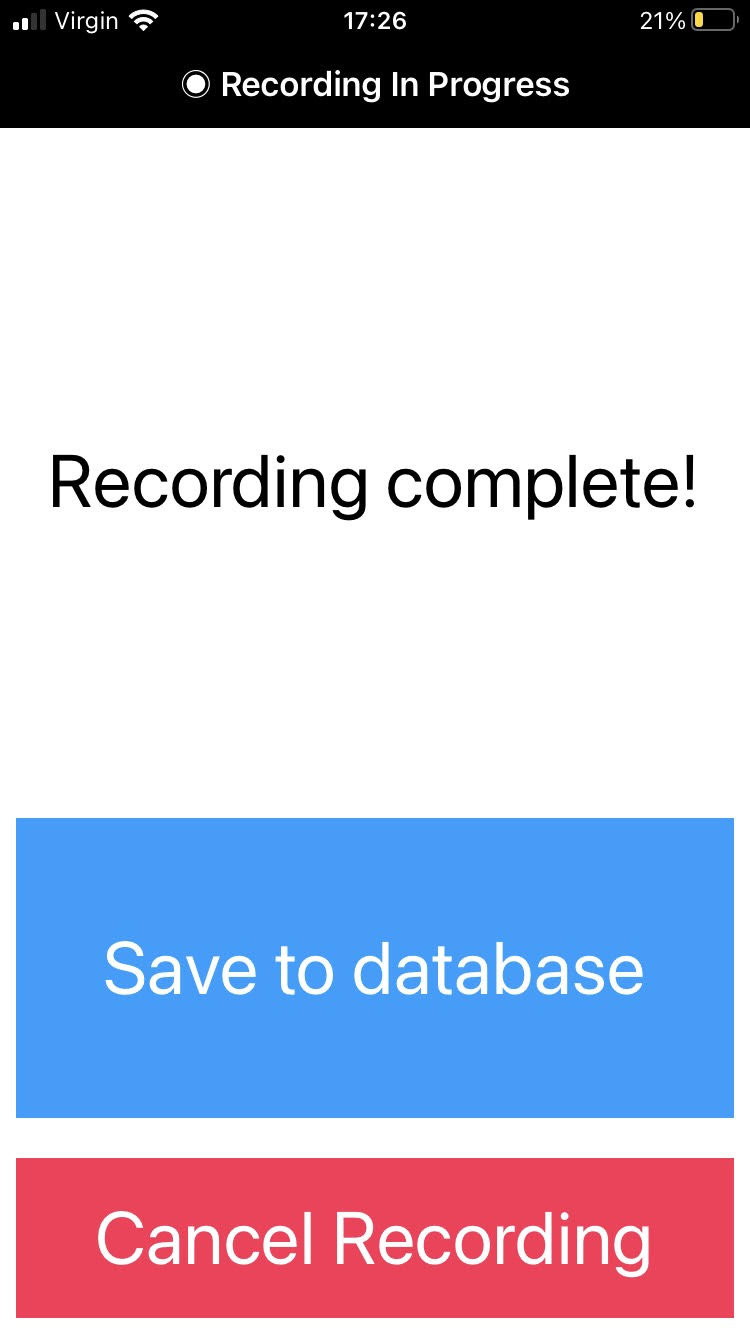}
\label{fig:lhfddg-recordingstopped}}
\quad
\subfigure[Live sensor outputs]{%
\includegraphics[width=0.15\textwidth]{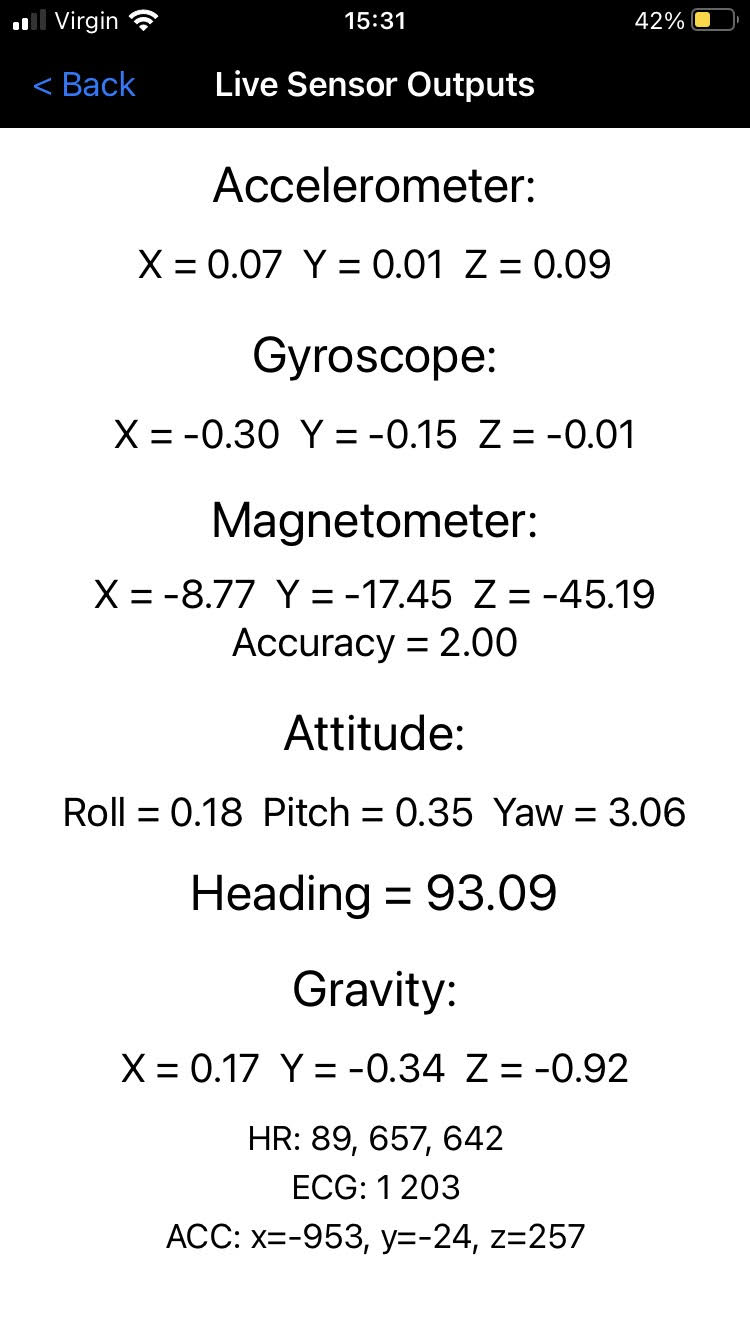}
\label{fig:lhfddg-livesensors}}
\quad
\subfigure[Settings]{%
\includegraphics[width=0.15\textwidth]{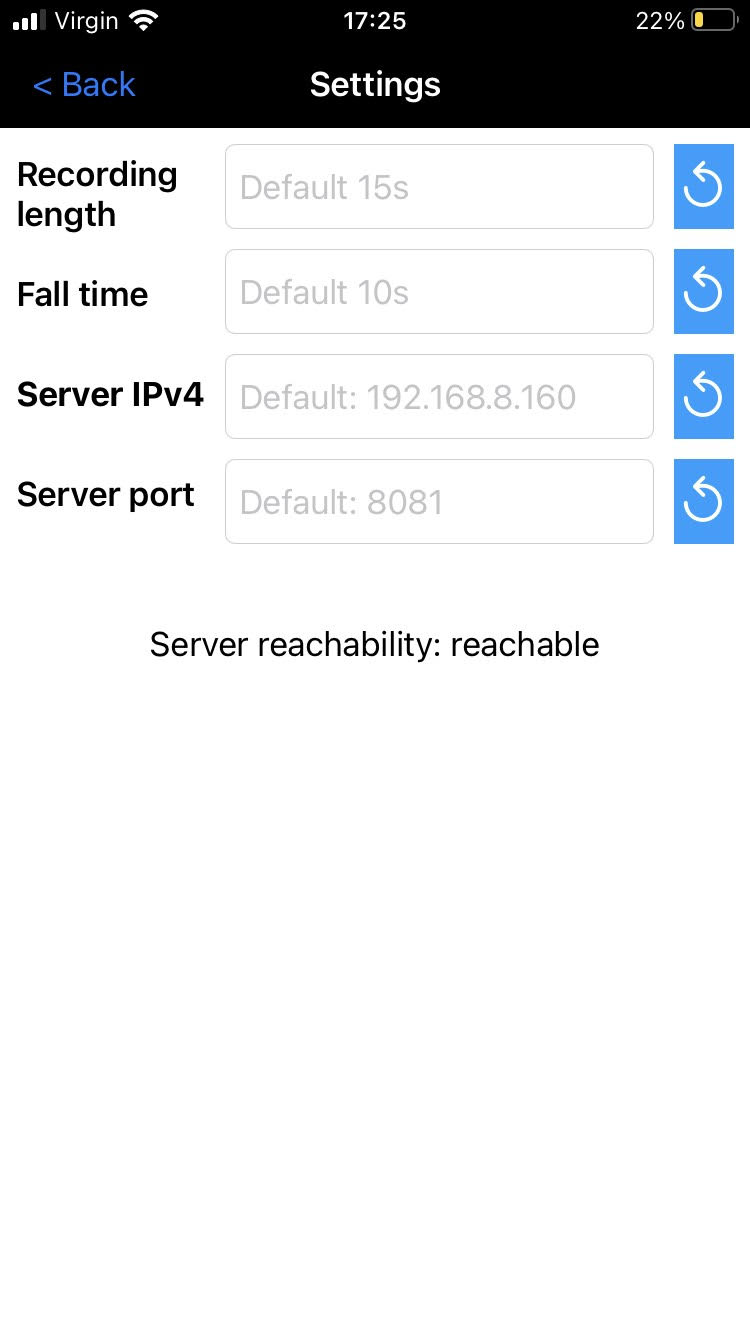}
\label{fig:lhfddg-settings}}
\caption{Other app pages}
\label{fig:figure1}
\end{figure}

\begin{itemize}
    \item \textbf{Connection} (fig \ref{fig:lhfddg-connection}): allows the user to connect/disconnect to the Polar H10 device via Bluetooth.
    \item \textbf{Add subject} (fig \ref{fig:lhfddg-addsubject}): allows the user to register a subject.
    \item \textbf{Create a recording} (fig \ref{fig:lhfddg-createrecording}): allows the user to create a recording if the provided subject ID can be found in my database (to ensure no incorrect annotations).
    \begin{itemize}
        \item \textbf{Recording in progress} (fig \ref{fig:lhfddg-recordingprogress}): plays sounds to signal subjects to fall over or get up, displays live sensor outputs so users can verify all the sensors are working as expected, a timer to show the duration of the recording, and a stop button to allow subjects to stop the recording.
        \item \textbf{Recording stopped} (fig \ref{fig:lhfddg-recordingstopped}): once the user stops a recording they can either choose to cancel the recording or save it to the database. If the user chooses to cancel the recording the recording metadata will not be saved to the database and so these orphan chunks will be filtered out in the preprocessing stage. If the user chooses to save the recording the recording metadata will be saved to the database along with any leftover chunks that failed to get posted during the recording.
    \end{itemize}
    \item \textbf{Menu} (fig \ref{fig:lhfddg-menu}): allows the user to navigate to 3 different pages.
    \begin{itemize}
        \item \textbf{Live sensor outputs} (fig \ref{fig:lhfddg-livesensors}): allows the user to see the real time outputs of the movement and ECG sensors. This was made as an easy way to verify the sensors were working as expected.
        \item \textbf{About}: provides a brief app description
        \item \textbf{Settings} (fig \ref{fig:lhfddg-settings}): allows the user to change the localhost server IP and port in case the IPv4 address of my computer changes when changing networks. Once this server address is changed a ping request is posted to the new address to check if it can make a connection, the outcome of this ping request is displayed in a "Server reachability" label so I can double check the connection is working after a network change (ie. when going to a lab on campus for data collection).
    \end{itemize}
\end{itemize}

\subsection{Storing Data}
The initial plan was to store all data on Firebase's servers using Firestore from \href{https://github.com/firebase/firebase-ios-sdk}{Firebase's Swift SDK} as this would facilitate the ability to store large amounts of data from my app online quickly and securely, and query/view this data in Firestore's online database management console. However, in order to obtain ethics approval for my study I had to consider the potential privacy implications and consequences of how I stored my data and thus hosting all my data on an external server was not viable given I was using biometric data.

\subsubsection{Localhost Server}
To overcome this issue a custom (localhost) server was implemented in NodeJS in order to connect the iOS data collection app to a database for data storage. This required creating an array of different request routes for getting, querying and posting Meta, Chunk, or User objects. A Swift networking SDK called Alamofire was then used to create server request methods from the iOS data collection app allowing further ease in performing these get, post, and query tasks.

\subsubsection{MongoDB Database}
A MongoDB database was used for this project due to it's high performance, simplicity, and good libraries in JavaScript (where I chose to run my server).

Classified as a NoSQL database program, MongoDB uses JSON-like documents with optional schemas. To get this working a standalone MongoDB instance was deployed on a PC, and the custom server was updated with the details for the MongoDB connection (this connection was setup using a package called Mongoose in JavaScript). Once this was done and the server script had been executed the MongoDB instance automatically connected to the server allowing for data storage.


\subsubsection{Circular Buffer}
The initial plan was to do fixed size recordings (15s) and send each entire recording to my server in a single request. After further testing and thought it was decided that it would be far more useful to have non-fixed recording length as this would allow me to collect data far more easily and quickly by performing longer recordings. However, this came with the limitation of the amount data that my phone could store at a given time without consuming too many hardware resources. Thus I decided to make use of a circular buffer which effectively processes the data in chunks, adds them to a post queue, and posts the chunks in the queue iteratively. At the end of the recording the metadata for the recording is then sent to the server which stores the IDs and order of all the relevant chunks to be merged. This alternative also provides further security against losing data as if a chunk fails to get posted that chunk will just get sent to the back of the post queue. Any unsent chunks at the end of a recording get sent along with the recording metadata.

\begin{figure}[ht]
\centering
\includegraphics[width=0.7\textwidth]{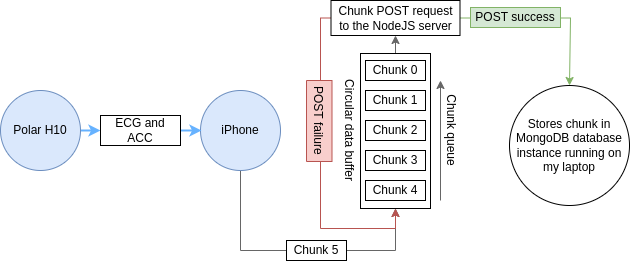}
\caption{Circular buffer data chunking protocol}
\label{fig:figure1}
\end{figure}

\subsection{Security \& Privacy}
The app will only be installed on my smartphone as I will always be the one conducting data collection. This takes away the risk of malicious users who could read/write data as it would be impossible for them to gain access to the app without breaking into my smartphone.

In order to ensure my localhost server was secure I added an IP whitelisting middleware to ensure that the server only processed requests from authorised devices (such as my laptop and phone). Although it is arguable that someone could spoof their IP and retrieve all my data it is extremely unlikely given this user would first need to be signed into the same local network as me (the University network), sniff network traffic to determine my server IP and port, and retrieve valid request routes for my server.

\begin{figure}[ht]
\centering
\includegraphics[width=0.55\textwidth]{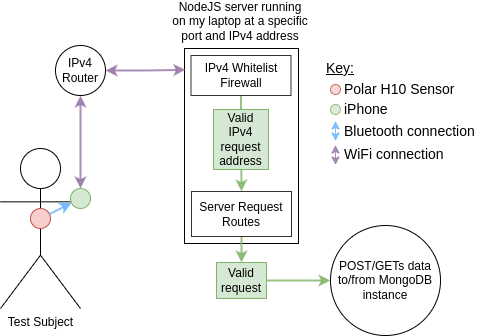}
\caption{Data transmission overview}
\label{fig:figure1}
\end{figure}

\section{Means of Data Collection}
This details the different considerations that had to be made when planning the data collection.

\subsection{Sensor Placement}
The iPhone was placed in the user's trouser pocket throughout the entirety of the recordings.

The Polar H10 device was worn snugly around the chest of the user under their shirt (to ensure constant skin contact for the ECG sensor electrodes). I thought this was particularly useful for fall detection as this device was placed just above a person's CG which should experience alot more acceleration than if it were on the waist when a fall occurs as this is further away from the ground. Furthermore, it is particularly useful due to the lack of noisy movements that part of the body experiences. For example, placing sensors on a watch would be alot more difficult due to the amount of different movements we perform with our arms (which are not related to falls).

\begin{figure}[ht]
\centering
\includegraphics[width=0.7\textwidth]{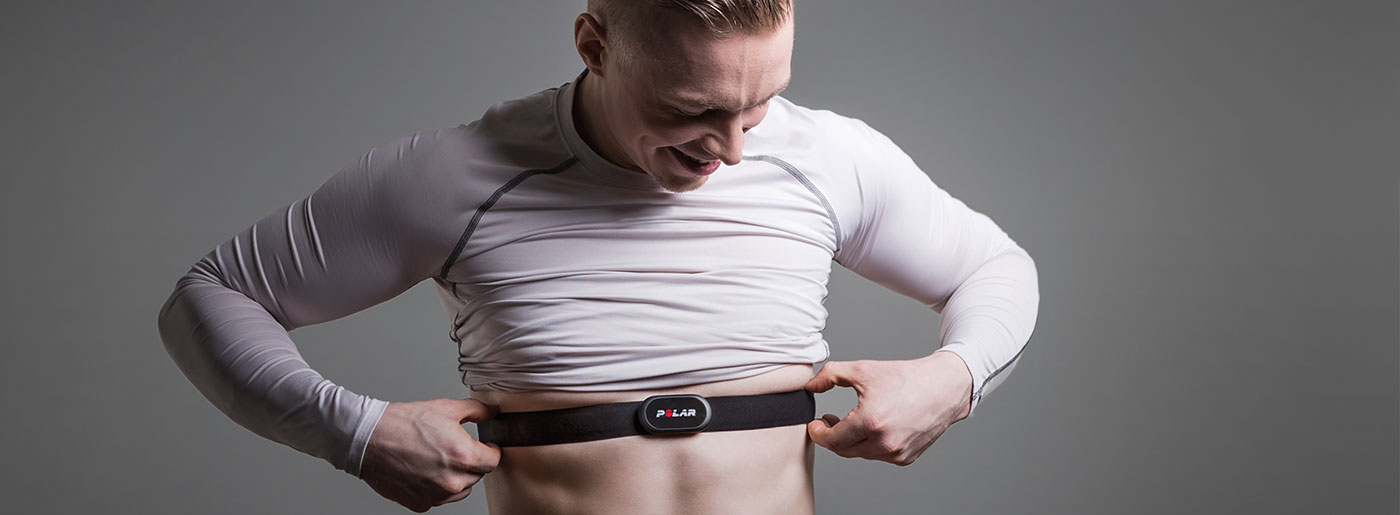}
\caption{Polar H10 chest strap placement}
\label{fig:figure1}
\end{figure}

\subsection{Data Recording Protocol}
Before we can start modelling this data using supervised learning techniques for fall detection we first need labelled data. Thus we must find an efficient and robust way this could be done. The most robust way to do this would be to have a human manually annotate each fall, however, this would add a lot of complexity and risk in the preprocessing stage as I would have to ensure I matched all of the labels to the correct data samples.

Thus it was decided that the most effective/efficient way to do it would be by playing sounds on the iPhone as a means to signal when the subject should fall and get back up. This would ensure that the labelling could all be done in the code of my app and thus would be robust, consistent, and easy. However, we must note this adds some lag between the initial labelling in my code and the actual action (fall/getting up) due to the fact humans have on average a 0.23s reaction time to auditory stimulus (\cite{jain_bansal_kumar_singh_2015}).

Throughout the duration of a recording I would tell the subject what type of fall to perform and kept talking to them so they were not focused on the fall in order to further promote the surprise factor for better generalisation in a real-world scenario.

\begin{figure}[ht]
\centering
\includegraphics[width=0.8\textwidth]{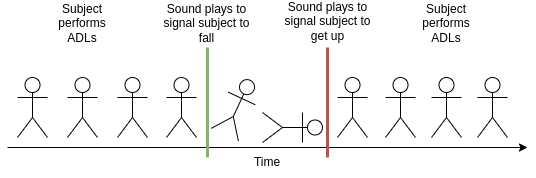}
\caption{Data recording protocol}
\label{fig:figure1}
\end{figure}

\subsection{Environment}
The data collection had to be performed in the University lab due to data privacy requirements. This was perfect for the experiment as it offered an array of objects to use for simulating falls/ADLs, however, it was a bit less ideal with regards to simulating the falls due to safety concerns. 

\begin{wrapfigure}{r}{0.3\linewidth}
  \begin{center}
    \includegraphics[angle=270,width=0.8\linewidth]{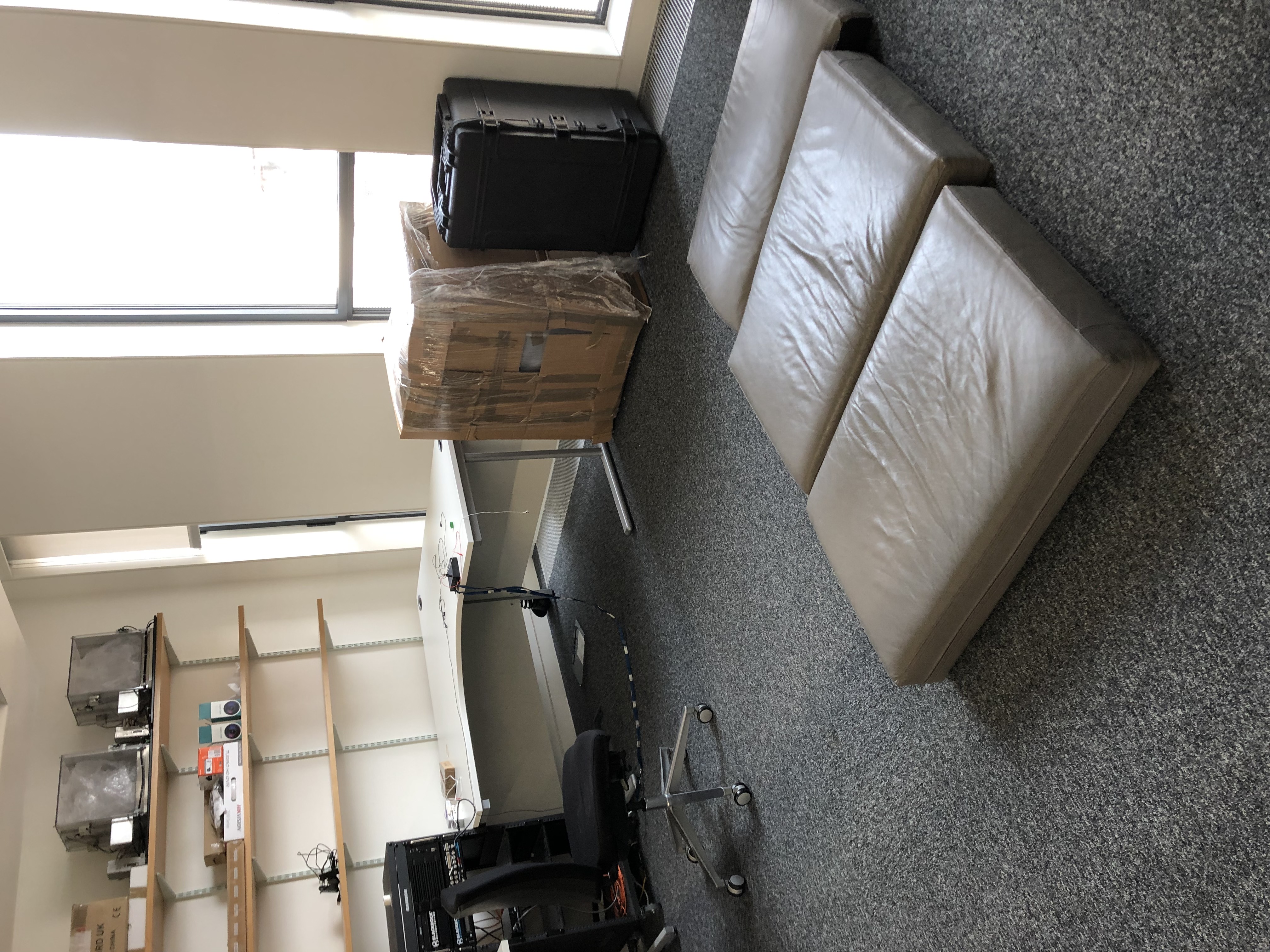}
  \end{center}
  \caption{Lab with mats setup for data collection}
\end{wrapfigure}

Multiple mats were placed on the floor in the center of the lab to ensure they were close enough for the test subjects to fall in various locations of the room while they were performing ADLs. Although, most of the time this did not pose any issues there were cases when test subjects were too far away and thus had to shuffle closer before initiating their fall. Any samples where this happened (which I could see would add a dangerous amount of labelling latency to the sample) were not saved to prevent skewing the training of the model. This is obviously not so efficient or safe and so it would be a lot more beneficial to use an environment with a fully padded floor allowing the test subjects to fall absolutely anywhere.

\subsection{Extra Safety Protocols}
Test subjects were provided the option to a wear a helmet in the study.

\section{Software Testing}

\subsection{Interfacing with Hardware Sensors}
A page on the iOS data collection app called 'Live Sensor Ouptuts' was used as a means to display live sensor outputs and thus see if these hardware sensors were working as expected (by seeing if they reacted to movement).

The main difficulty with this was ensuring the Polar and iPhone sensor outputs were aligned correctly given Polar outputs had an associated transmission latency from the Polar H10 chest strap.

Setting up the Polar sensors proved a lot more difficult than the iPhone's due to the initial setup required before I could even start performing the data transmision. This initially required me to set up BlueTooth connection handling functions to connect to the Polar device. Once this was set up, I tried to start requesting the data streams for the ECG and Accelerometer features, however, these kept failing. Thus I decided to post an issue on Polar's SDK repository on GitHub. This was quickly attended to, however, even the developer that responded to my ticket was unsure the cause of this issue. After long testing I finally realised it was due to the fact I was trying to initiate data streams immediately after a successful BlueTooth connection was made with the device. This was problematic as these data streams needed to be marked as ready first before they could be streamed from. Thus I adapted my code to only request data streams after these streams were marked as ready, which solved all the streaming issues.

\subsection{Data Transmission \& Storage}
In order to test my data collection system worked I demonstrated the app to my supervisor, Kianoush Nazarpour.

My initial demonstration was delayed due to a small fault in my iOS app which did not allow me to change server IPv4, and port addresses as the input keyboard was of the wrong type (a number keypad was used however IP addresses also contain full stops).

After fixing this issue everything worked as expected.

My supervisor then suggested to rather change my recordings from a fixed length (15s) to a variable length to make data collection more efficient and comparable to a real-life scenario. This added a significant layer of complexity to my work as storing such a large amount of data would be infeasible to do on the phone. 

I had to rather implement a circular buffer that would send recording chunks iteratively throughout the recording (every 5s). This also added some complexity if I wanted to cancel a recording as the chunks would have already been sent to the server. To handle this I decided that when a recording was saved a recording Meta object would be sent to the server. This Meta object would hold all the data needed to remerge the chunks of a given recording: ordered chunk indexes, user data (subject ID), and creation date (for debugging purposes).

\subsection{Circular Buffer}
Given the new addition of the circular buffer in the app for data transmission of variable length recordings, this had to be tested thoroughly before data collection was started.

I tested this on myself in my flat to ensure all data was sent.

This was difficult at first as at times chunks would fail to be sent in the post queue. Intuitively, one might think sending any failed chunks immediately after a post failure would solve this issue, however, this is not the case as we can easily run into infinite loops (if a post request that keeps failing keeps getting resent). Thus I decided to rather send any unsent chunks at the end of the recording if the recording was saved. This method proved to work perfectly.

\subsection{Data Exporting}
It would be dangerous to move to the data collection stage without performing a full system test and seeing if the exported data was valid.

Thus a couple of fall recordings were collected by the author in their flat and exported into bson files for processing.

\textbf{The following checks were performed on the resultant data:}\\
Given fear and surprise result in a raised heart rate it is important to be careful when interpreting the ECG results for our experiment. 

Due to the fact that in this experiment I told subjects to fall over it is likely the collected ECG is slightly different from a real-world fall due to a reduction in fear/surprise experienced by the subject. However, we must also note that cardiovascular stress can also result in raised ECG levels and thus is still representative.

To determine if the means of data collection resulted in any significant differences in the ECG results test recordings with simulated and non-simulated falls were conducted. These test recordings were then parsed using my preprocessing code in Python (discussed in the next chapter), and visualized each one to get an idea of the relation between sensor outputs and fall state over time.

\begin{figure}[ht]
\centering
\subfigure[Recording 1]{%
\includegraphics[width=0.25\textwidth]{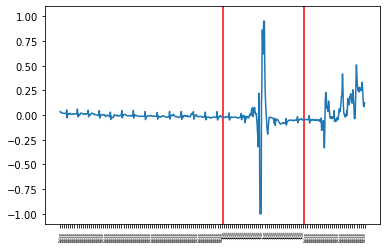}
\label{fig:lhfddg-recordingprogress}}
\quad
\subfigure[Recording 2]{%
\includegraphics[width=0.25\textwidth]{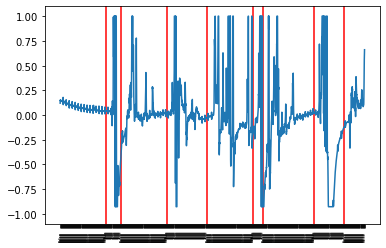}
\label{fig:lhfddg-settings}}
\caption{ECG data for simulated falls}
\label{fig:figure1}
\end{figure}

All these recordings seemed to indicate that were was a significant change in ECG during a fall. This is particularly exciting as ECG sensors have not been widely used as a means for fall detection but rather systems typically focus on using movement sensors such as accelerometers and gyroscopes. One might argue these results can be attributed to an electrode slip of the sensor, however, I do not believe this is the case given the strap has silicone dots to prevent the electrodes slipping and the strap was fitted snugly on each of the test subjects. 

\begin{figure}[ht]
\centering
\subfigure[Recording 1]{%
\includegraphics[width=0.25\textwidth]{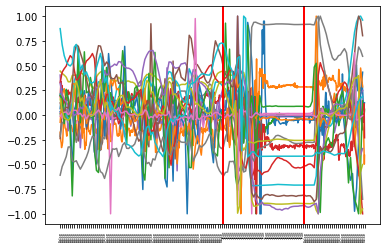}
\label{fig:lhfddg-recordingprogress}}
\quad
\subfigure[Recording 2]{%
\includegraphics[width=0.25\textwidth]{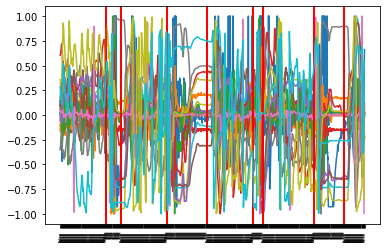}
\label{fig:lhfddg-settings}}
\caption{All sensor data for simulated falls}
\label{fig:figure1}
\end{figure}

All these sensor results help us visualize the stages of a fall more easily: the time for the subject to initiate the fall after hearing the sound, the time for the subject to fall, and the time for the subject to get up after lying on the ground.

\section{Resulting Dataset}

\subsection{Iterations}
To initially verify that the data collection application worked and that the resultant data was robust some test data was collected personally by the author (falls were simulated by falling onto cushions on the floor).

The final dataset was collected over two 3-4 hour sessions in the lab.

In the first session 3 subjects were recorded while performing a small set of ADLs: walking, standing, sitting, leaning. After this session I developed a machine learning model to test the data, this model was ported onto an iPhone and tested on myself. It worked perfectly except in the cases where I were to jump or go up stairs. Thus I decided I needed to expand my set of ADLs to cater for these cases.

In the second session 5 subjects were recorded (including the same 3 from the first session) while performing a few extra ADLs: tying shoelaces, walking and quickly stopping, walking with quick changes in direction, jumping over something (ie. a puddle), step up/down a step, jump onto something (ie. a step), and kneeling/leaning over to get something from a bag on the floor.

\subsection{Sampling Frequency}
The iPhone sensor data was sampled at 10Hz, the Polar ECG sensor data was sampled at 130Hz, and the Polar accelerometer sensor data was sampled at 200Hz. The Polar sensor data was streamed at maximum possible frequencies to ensure this data was as detailed as possible. The frequency was limited to 10Hz for the iPhone sensors due to hardware limitations given the large array of sensors used, and the fact these sensors were of lesser importance to the Polar sensors.

\subsection{Sample Representation}

This data collection recorded 5 test subjects with an average of 13.4 recordings per subject, an average recording duration of 117s, and an average of 61 falls per subject. The average subject demographic was a 22 year old male at an average height of 176.2cm, and an average weight of 74.4kg.

\subsubsection{Test Subjects}
Given the majority of the test subjects were young and healthy University students the dataset is less representative of the elderly population. However, this was mainly due to the difficulty and danger of using elderly subjects in such an experiment.


\subsubsection{Minimising Subject Bias}
In order to minimise bias in the way a subject chose to simulate their falls it was ensured that each participant was given the exact same breakdown as to what to do.

\subsubsection{Falls \& ADLs Performed}
To ensure the data was representative enough to be used for fall detection a variety of ADL, and fall types were used to ensure my model was able to distinguish between the two.

I used the fall types found from my domain specific research: stumbling, slipping, fainting, falling from a high structure, and falling after a jump. I also asked participants to perform these falls in different directions (where appropriate) to account for as many fall scenarios as possible.

I used some typical ADLs based off existing fall datasets: walking, going up a step, going down a step, sitting on a chair, getting up from a chair. As well as, a couple of custom examples that I decided would also be useful: tying shoelaces, walking and quickly stopping, walking with quick changes in direction, jumping over something (ie. a puddle), and kneeling/leaning over to get something from a bag on the floor.

Throughout each session with my test subjects I also asked them what actions they felt they had not done so much and what other actions they felt should be represented. I thought this was particularly useful as it allowed participants to add extra actions that other participant's might not do, allowing my data to be more realistic.

\chapter{Data Preprocessing}

\section{JSON Parsing}
Raw BSON data was parsed from the MongoDB database using the \href{https://pymongo.readthedocs.io/en/stable/api/bson/index.html}{bson} Python library. This library allowed JSON documents to be parsed from all the different schemas (Meta, Chunk, and User) into 3 separate dictionaries (a Python data structure) with the document IDs used for the keys. Custom classes were then created for each of the schemas in which these dictionary elements were used to initialize each object.

Next, a custom Recording class was created in order to store all the data for a given recording. This class would retrieve all the Chunk objects with a specified recording ID, order them by chunk index, and append their data to the given Recording class object. Thus parsing all the recordings could be done by just iterating over the Meta objects parsed earlier, and initialize corresponding Recording objects for each of these.

\section{Preprocessing Evaluation}
In order to determine which preprocessing techniques would be the most useful the preprocessed datasets were evaluated against simple SciKit-Learn baselines (ie. Linear Regression). These results were not included for the sake of brevity.

\section{Feature Selection}
For the data collection stage I thought it would be the most useful to record as much data as possible and when the time came for model building I could then just use optimisation techniques to choose the best subset of these features. 

Removing these unnecessary features is important as it enables us to train our model quicker (particularly useful when updating model fits to new data), reduce the complexity of our model, reduce overfitting, and could even improve the accuracy of our model if the right subset is chosen (\cite{fselection_2021}). These small improvements in speed and efficiency are all particularly important for such a time-sensitive use-case like fall detection.

In this case, using the iPhone sensors would be detrimental given that all samples were collected with this phone in a static position (the subject's trouser pocket). This is evidently not representative of a true scenario where a user may hold their phone, put it in their bag, drop it etc. Thus it was decided to exclude all iPhone sensor data from the datasets.

\section{Feature Shifting and Scaling}
There are two main techniques used for feature shifting and scaling in machine learning, namely normalisation and standardisation (\cite{feature_scaling_2020}). The choice between these two techniques is fully dependent on the context and distribution of data that we will be using.

\textbf{Normalisation:} $X' = \frac{X - X_{min}}{X_{max} - X_{min}}$\\
\textbf{Standardisation:} $X' = \frac{X - \mu}{\sigma}$

Due to the importance of outliers in the context of fall detection (we can treat falls as an outlier event) I anticipated standardisation would be the most useful transformation due to its robustness against outliers (\cite{outlier_scaling_2020}). However, for the sake of optimisation I tested both techniques against my dataset and chose the one which produced the more optimal model.

As expected, standardisation proved more effective than normalisation. This could indicate that this sensor data may follow Gaussian distribution's as unlike normalisation (which does not assume any data distribution) standardisation is typically used on Gaussian distributed data.

Log transform was also evaluated on strictly positive features (ie. age, height, weight) that may result in right-skewed distributions, however, after evaluation pure standardisation still proved more effective and thus this transformation was not used.

\section{Time Series Data Techniques}

\subsection{Lag Features}
Lag features are a clever way to manipulate data so that we can predict an event in the future. This is evidently very useful for an application such as fall detection where predicting a fall before it is too late could pose massive benefits. If such a model were to be produced it could be used to trigger safety devices such as an airbag to mitigate the likelihood of injury. Lag features work by shifting the labels from the data they correspond to, to data in the past. This would ultimately allow us to fit a model that predicts whether or not a fall will occur or not X milliseconds in the future.

Multiple models were created with varying values of X to see how far we could push the accuracy of our model for fall predictions. Due to the fact a fall is initiated within a 0.5s window testing was only performed on values of X from the range 0 to 500 in steps of 100.


\subsection{Sliding Windows}
Given a sequence of data for a time series dataset, we can restructure the data to look like a supervised learning problem. We can do this by using all the data for a given window (or time frame) as inputs. Thus we must choose a sensible window size to ensure sufficient detail for classification whilst still maintaining a relatively low algorithmic complexity. Given falls are a fast event (around 0.5s of fall time) I would not expect that we would have to use such large window sizes. 

Given this domain knowledge I decided to use 1-2s window sizes (10-20 recording intervals of 100ms) for fall detection and prevention.

\section{Labelling}\label{sec:labelling}
The number 1 was used to represent a fall, and 0 to represent no fall. Given we decided to use sliding windows, a single sample of $w$ 100ms data intervals (where $w$ is the window size) will also have $w$ labels when it only needs 1. We will have to find a meaningful way to process this in order to maximize performance for ML models. A sample was labelled as a fall if any of the intervals had a fall label. For reference I decided to call this technique "Existence labelling". This process is shown formally below where $\Vec{Y}$ is the vector of interval labels, and $\hat{y}$ is the resultant label for those intervals:

\begin{equation}
  \hat{y}=\left\{
  \begin{array}{@{}ll@{}}
    1, & \text{if}\ {\sum_{i=1}^{w} y_i > 0}\\ 
    0, & \text{otherwise}
  \end{array}\right.
\end{equation} 

\begin{center}
     where $\Vec{Y} = \{y_1, ..., y_w\}$
\end{center}

\section{Time Domain VS. Frequency Domain}
When working with signal data we can choose to use FFT/DFT\footnote{Fast Fourier Transform or Discrete Fourier Transform} for transforming the data from the time domain to the frequency domain. In the time domain, a signal is a wave that varies in amplitude (y-axis) over time (x-axis). However, in the frequency domain a signal is represented as a series of frequencies (x-axis) that each have an associated power (y-axis) (\cite{signal_proc_fft}). Given the complexity of adding this as another possible preprocessing technique it will not be evaluated in this paper.

\section{Resulting Datasets}
Multiple datasets were created with varying preprocessing techniques (described below) in order to evaluate which would perform best. This preprocessing variation included varying the window size, meaning the size of my datasets varied with window size (due to the usage of overlapping windows) and thus the number of samples across the datasets was not constant.

The datasets have on average 62188 samples (across window sizes 10-20) with $75w$ features per sample, and an average fall to ADL sample ratio of 22:78 (meaning on average 22\% of samples are labelled as falls). All the features were standardised before being randomly shuffled and split into training, validation, and testing sets (of size 60\%, 20\%, 20\% respectively). Existence labelling (described in section \ref{sec:labelling}) was used in all datasets.

\subsection{Sample description}
Samples were of size $(w,75)$ (for sliding windows of size $w$) which consisted of $w$ 100ms intervals of 75 features with 2 features for user height and weight.

\subsubsection{The 75 features per 100ms interval:}
\begin{itemize}
    \item Polar ECG (size 13)
    \item Polar Accelerometer XYZ (size 20*3)
    \item User height (cm) (size 1)
    \item User weight (kg) (size 1)
\end{itemize}

It would have been preferred to not have had to include the user height and weight in each 100ms sample for efficiency, however, shaping the data in a meaningful way prevented me from doing so. This is mainly due to the fact I wanted to use a dimension to represent the window size $w$, and another to represent the 73 Polar features for a given 100ms interval. This data configuration ultimately aligns data from the same sensors across adjacent windows (resulting in a $(w,73)$ sample size) which is particularly important for CNN models where spatial information is just as important as the feature values (given this architecture is used for images) (\cite{tang_long_liu_zhou_jiang_blumenstein_2020}).

Evidently given this sample shape $(w,73)$ it would be impossible to simply just append the two user features on the end. I thought about parsing these features separately (the sensor and user data) by using a nested architecture (a model with nested models that parse the tabular and image data separately), however, I decided it would be more useful for the sake of model complexity to rather just add these user features to each interval (resulting in $(w,75)$ size samples).


\subsection{Preprocessing Variability}
The sample window size $w$ will be varied in the modelling experiments in order to evaluate ML fall detection models of varying complexities.

\underline{Sample window sizes used:} $w \in \{10, 20\}$

\chapter{Fall Detection Modelling}

\definecolor{agreen}{rgb}{0.55, 0.71, 0.0}
\definecolor{cgreen}{rgb}{0.0, 0.8, 0.6}
\definecolor{dgreen}{rgb}{0.05, 0.5, 0.06}

\section{Evaluating Model Performance}
Given falls occur less frequently than ADLs in a realistic scenario, there is a class imbalance in my data. When fitting data with class imbalances it is extremely important to use appropriate metrics as otherwise it is possible to interpret your model performance as being better than it truly is. For example, if our data had a class imbalance of 0.15 (meaning 15\% of samples were a fall) then our model could easily achieve 85\% accuracy by just classifying every sample as not a fall.

Thus throughout all experiments AUC\footnote{The area under the ROC curve. This metric ultimately tells us how much the model is capable of distinguishing classes.}, sensitivity (true positive rate), and specificity (true negative rate) were used to evaluate the models.

Given this is for a safety-critical application we want high sensitivity scores so that if there is a fall it is unlikely the model would not recognise it.

\underline{Please note in all the results tables:}
\begin{itemize}
    \item "sens." denotes sensitivity, and "spec." denotes specificity
    \item The best results of a metric (ie. Test AUC) for a given window size (ie. 10) are marked in \textbf{bold}, the best overall results for each metric over all window sizes are marked in \textcolor{red}{\textbf{red}}
    \item For a given model, the best model performance throughout the training epochs was used (determined by the epoch which produced maximum validation AUC)
    \item All the deep learning models were trained over 100 epochs with an SGD optimizer (learning rate 0.01, weight decay 1e-4, momentum 0.9), a learning rate scheduler (which reduces learning rate on plateau with factor 0.1 and patience 5), with a batch size of 512.
\end{itemize}

\section{Baseline Models}
Simple baseline models in SciKit-Learn were tested on the datasets first. This is extremely important to gauge the quality of data and to gain a point of reference when developing more complex models. When developing complex models, bugs in the code (that hinder model performance) are much easier to identify as we would know that the model should be performing significantly better than the baseline results.

\begin{table*}[ht]
    \hspace*{1cm}
    \scalebox{0.9}{
    \begin{tabular}{l|c|cc|cc|cc}
    \toprule
        Model &  Window & Val. & Test & Val. & Test & Val & Test \\
         & size & AUC & AUC & sens. & sens. & spec. & spec. \\
    \midrule
        Linear Regression   & 10   & 57.81 & 57.34 & 18.63 & 17.89 & 96.99 & 96.78 \\
        Decision Tree       & 10   & 66.65  & 65.11 & 48.34 & 44.99 & 84.96 & 85.24 \\
        SVM                 & 10   & 62.11 & 60.82 & 26.48 & 24.04 & \textcolor{red}{\textbf{97.74}} & \textcolor{red}{\textbf{97.60}} \\
        KNN (k=3)           & 10   & 67.29 & 65.59 & 40.65 & 37.42 & 93.93 & 93.75 \\
        MLP (1x500)         & 10   & 68.91 & 66.98 & 49.17 & 45.96 & 88.65 & 88.00 \\
        Bernoulli NB        & 10   & 66.48 & 64.31 & \textbf{57.58} & \textbf{53.89} & 75.39 & 74.73 \\
        Gaussian NB         & 10   & \textbf{69.35} & \textbf{67.11} & 52.21 & 48.01 & 86.49 & 86.21 \\
    \midrule
        Linear Regression   & 20   & 63.26 & 63.56 & 29.60 & 30.22 & \textbf{96.93} & \textbf{96.90} \\
        Decision Tree       & 20   & 71.33 & 70.62 & 57.67 & 56.78 & 85.00 & 84.46 \\
        SVM                 & 20   & 70.96 & 71.42 & 45.77 & 46.89 & 96.15 & 95.94 \\
        KNN (k=3)           & 20   & \textcolor{red}{\textbf{72.54}} & \textcolor{red}{\textbf{72.17}} & 51.18 & 50.86 & 93.91 & 93.48 \\
        MLP (1x500)         & 20   & 68.53 & 68.90 & 49.29 & 50.53 & 87.76 & 87.27 \\
        Bernoulli NB        & 20   & 67.90 & 67.46 & \textcolor{red}{\textbf{59.20}} & \textcolor{red}{\textbf{58.43}} & 76.61 & 76.49 \\
        Gaussian NB         & 20   & 72.20 & 71.38 & 58.30 & 57.67 & 86.10 & 85.09 \\
    \bottomrule
    \end{tabular}}
    \caption{Experiment results of multiple different baseline ML models for varying window sizes and 0ms labelling lag. These models were all developed using the SciKit-Learn ML library in Python. Keywords: SVM (Support Vector Machine), KNN (K-Nearest Neighbours), MLP (Multi-Layer Perceptron), NB (Naive Bayes).}
    \label{tab:baseline_results}
\end{table*}

\section{Deep Learning Models}
Deep learning models refer to a subset of neural networks with 3 or more layers (including the input and output layers). These types of models have been found to provide great performance on datasets due to their lack of assumptions about the data, and ability to develop powerful features for classification.

\subsection{Result Consistency}
In order to keep results comparable from varying models the same random seed was used in all cases to ensure the model weights were initialised in a consistent way.

\subsection{Preventing Overfitting}
The following techniques were used in all experiments as a means to prevent overfitting and promote generalisation.
\begin{itemize}
    \item \underline{Parameter normalisation penalties:} L1/L2 regularization. This adds an extra term to the weight update of the model, helping make the model more robust against outliers and noise.
    \item \underline{Early stopping:} simply stopping the model at an optimal point (before it starts to overfit). However, validation and test scores were tracked across every model iteration and thus we were able to calculate the optimal stopping point after the full training cycle for each model.
\end{itemize}

\subsection{Long Short-Term Memory Networks}
Long short-term memory networks are an artificial recurrent neural network (RNN) architecture. This architecture was specifically designed to employ long short-term units and gated recurrent units as a means to process sequences (\cite{luna_2019}). Recent studies have also shed some light on the potential for dynamic signal classifications, and accelerometer data implying the relevance of such an architecture for fall detection.

\subsubsection{Experiments}
The depth and width of LSTM networks were varied across the experiments as a means to vary model complexity (and thus ability to generalise):

\begin{table*}[ht]
    \hspace*{0.4cm}
    \scalebox{0.9}{
    \begin{tabular}{l|c|c|cc|cc|cc}
    \toprule
        Model & Window & Optimal & Val. & Test & Val. & Test & Val & Test \\
         & size & epoch & AUC & AUC & sens. & sens. & spec. & spec. \\
    \midrule
        LSTM (2x400) & 10   & 97 & 67.78 & 66.48 & 38.63 & 36.19 & 96.92 & 96.76 \\
        LSTM (4x400) & 10   & 94 & 67.98 & 67.04 & 39.86 & 38.01 & 96.10 & 96.07 \\
        LSTM (2x600) & 10   & 95 & 68.87 & \textbf{68.14} & 40.81 & 39.40 & \textbf{96.93} & \textcolor{red}{\textbf{96.88}} \\
        LSTM (4x600) & 10   & 98 & \textbf{69.28} & 68.04 & \textbf{42.66} & \textbf{40.37} & 95.90 & 95.71 \\
    \midrule
        LSTM (4x400) & 20   & 99 & 74.17 & 74.00 & 51.31 & 51.36 & \textcolor{red}{\textbf{97.02}} & \textbf{96.64} \\
        LSTM (6x400) & 20   & 71 & 73.11 & 72.44 & 50.25 & 49.44 & 95.96 & 95.45 \\
        LSTM (4x600) & 20   & 69 & \textcolor{red}{\textbf{75.66}} & \textcolor{red}{\textbf{75.80}} & \textcolor{red}{\textbf{54.92}} & \textcolor{red}{\textbf{55.02}} & 96.40 & 96.57 \\
        LSTM (6x600) & 20   & 68 & 74.02 & 73.52 & 51.91 & 51.36 & 96.14 & 95.67 \\
    \bottomrule
    \end{tabular}}
    \caption{Experimental results of LSTM models for varying window sizes, layer depths, and layer widths.}
    \label{tab:lstm_results}
\end{table*}

\subsection{Convolutional Neural Networks}
The Convolutional Neural Network architecture was initially developed as a method for image classification. Thus this architecture accepts a sample image (a matrix of size MxN) and performs feature extraction and classification via hidden layers (such as convolutional layers, RELU layer, max-pooling layers). However, it has also been recognised as a useful architecture even for non-image data due to the way CNNs use locational information (\cite{tang_long_liu_zhou_jiang_blumenstein_2020}). Unlike traditional neural networks CNNs information is not extracted from a feature's value but rather localised patterns denoted by this feature's value in relation to other close-by features (pixels). We can see that this also applies a lot to time-series data given feature values are only really useful for identifying complex actions when given the adjacent recording interval feature values.

This model is particularly favoured in the field of machine learning given the flexibility of model due its minimal memory footprint, and robust feature extraction (\cite{sharma_vans_shigemizu_boroevich_tsunoda_2019}):
\begin{itemize}
    \item It does not require additional feature extraction as it automatically derives features from the raw elements
    \item It finds higher-order statistics of image and nonlinear correlations
    \item Convolution neurons process data for its receptive fields or restricted subarea. relaxing the need to have a very high number of neurons for large input sizes and therefore enables the network to be much deeper with fewer parameters
    \item It uses weight sharing (ie. many receptive fields share the same weights, biases, and filters) enabling a reduction in the memory footprint as compared to conventional neural networks
\end{itemize}

Evidently, CNNs provide many benefits as a ML model, however, in order to use it on our non-image data we must reshape our samples from 2D into 3D matrices. This additional dimension is required in order to mimic the input channels of an image (3 for a coloured RGB image, and 1 for a grayscale image).

However, we must bear in mind that the positioning of respective pixels can adversely affect the feature extraction and classification performance of CNN architecture if arbitrarily arranged. It was decided to test varying input positions as a means to find the configuration that produces the best validation accuracy.

A test was performed using inputs of size $(w,1,75)$ which sets the number of input channels to be the window size, and inputs of size $(1,w,75)$ which sets the the number of input channels to be 1. However, in all cases using inputs of size $(1,w,75)$ proved more effective and so these sample sizes were used for all CNN models.

\subsubsection{Experiments}
Given the ResNet (a CNN-based architecture) architecture achieved state-of-the-art results for wearable-based fall detection (\cite{zhang_li_wang_2021}) this architecture was used across all CNN experiments with varying depths (ie. ResNet18 implies a depth of 18) as a means to vary model complexity (and thus ability to generalise). The specific ResNet architecture I used was developed by \cite{he_zhang_ren_sun_2015} due to it's 112466 citations, and given it won 1st place in the 2015 ILSVRC 2015 classification task.

\begin{table*}[ht]
    \hspace*{0.4cm}
    \scalebox{0.9}{
    \begin{tabular}{l|c|c|cc|cc|cc}
    \toprule
        Model & Window & Optimal & Val. & Test & Val. & Test & Val & Test \\
         & size & epoch & AUC & AUC & sens. & sens. & spec. & spec. \\
    \midrule
        ResNet18    & 10   & 37 & 77.29 & 75.70 & 61.92 & 58.70 & 92.66 & 92.69  \\
        ResNet34    & 10   & 43 & 77.46 & 76.18 & 61.21 & 58.59 & 93.71 & 93.77 \\
        ResNet50    & 10   & 26 &  76.94 & 75.18 & 60.06 & 56.99 & 93.82 & 93.37  \\
        ResNet101   & 10   & 33 &  \textbf{79.03}  & \textbf{77.64}  & \textbf{62.75} & \textbf{60.01} & \textbf{95.32} &  \textbf{95.28} \\
        ResNet152   & 10   & 46 & 77.86 & 76.62 & 60.69 & 58.78 & 95.02 & 94.47 \\
    \midrule
        ResNet18    & 20   & 48 & 89.01 & 89.39 & 80.84 & 81.32 & 97.17 & 97.46 \\
        ResNet34    & 20   & 52 & 90.34 & 90.59 & 82.90 & 83.33 & 97.78 & 97.84 \\
        ResNet50    & 20   & 43 & 90.72 & 90.91 & 83.76 & 84.06 & 97.69 & 97.75 \\
        ResNet101   & 20   & 42 & 90.82 & 91.55 & 83.43 & 84.85 & 98.21 & 98.24 \\
        ResNet152   & 20   & 64 & \textcolor{red}{\textbf{92.72}} & \textcolor{red}{\textbf{92.80}} & \textcolor{red}{\textbf{87.24}} & \textcolor{red}{\textbf{87.27}} & \textcolor{red}{\textbf{98.20}} & \textcolor{red}{\textbf{98.33}} \\
    \bottomrule
    \end{tabular}}
    \caption{Experimental results of ResNet models for varying window sizes, and layer depths.}
    \label{tab:cnn_results}
\end{table*}

\section{Fall Prevention}
Fall prevention was tested on the best performing model for 100ms and 200ms labelling lag. However, in all cases only very small progressions in AUC were made (the maximum val. AUC found was 51.58\%) and thus these results were not included.

\section{Evaluation}

\begin{figure}[ht]
\centering
\includegraphics[width=0.7\textwidth]{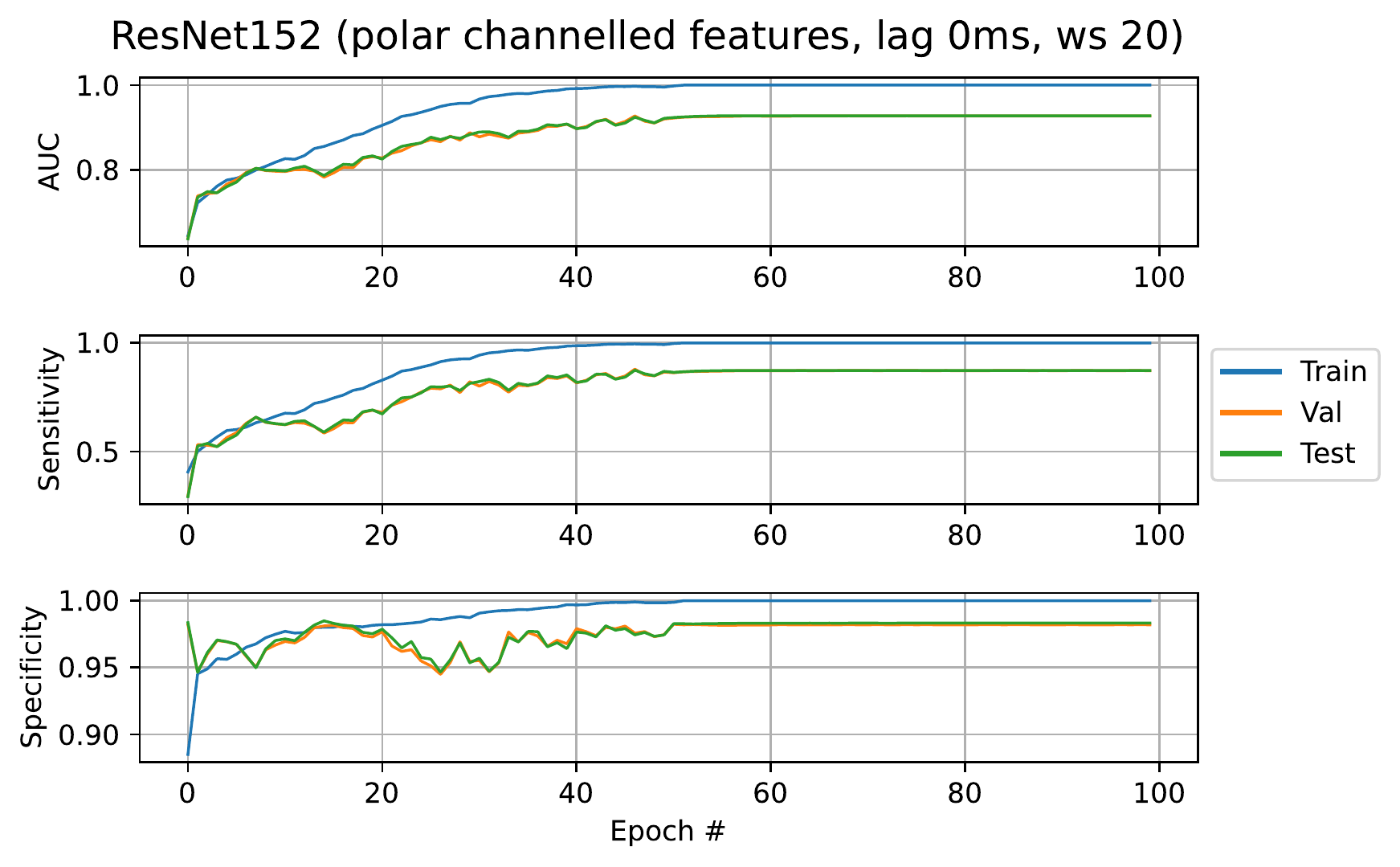}
\label{fig:lhfddg-recordingprogress}
\caption{A graph to show the evaluation metrics of the best performing model (ResNet152 with window size 20) over all epochs of the training cycle for the train, validation, and test sets.}
\end{figure}

\section{Model Exporting}
In order for the fall detection model to be efficiently run on a mobile phone it needs to be exported into a tflite model as this allows for optimised on-device machine learning which addresses latency, privacy, connectivity, size, and power consumption (\cite{tflite_2022}).

\ul{Given all models were developed in PyTorch the model exportation had to go through the following conversions:}\\
PyTorch $\rightarrow$ ONNX $\rightarrow$ TensorFlow $\rightarrow$ TensorFlowLite

\chapter{iOS Fall Detection System}

A final full system implementation was developed as a means to evaluate/verify the performance of my various fall detection models on live data, and provide a prototype commercial app.

\section{User Interface: iOS app}

\subsection{Functionalities}

\subsubsection{Commercial Functionalities}
\begin{itemize}
    \item Connected to a secure NoSQL database in the cloud which allows for:
    \begin{itemize}
        \item Fast data read/write speeds
        \item Setting user read/write privileges for privacy/security
        \item Create account functionality to allow users to store personal information (ie. emergency contacts, and usage statistics)
        \item Log in functionality to allow users to access their account
        \item Forgot password functionality to allow users to reset their account password if forgotten via a password reset email link
    \end{itemize}
    \item Live plots of sensor outputs: Polar accelerometer xyz, and Polar ECG
    \item Biometric and fall data statistics: \# falls detected, model TPR, model FPR, average heart rate etc.
    \item Detect falls in real time when the live fall detector is turned on
    \item Runs the live fall detection model on background threads for computational efficiency
    \item Sends a notification to the user if a fall is detected
    \item Texts emergency contacts if a fall is detected and the user is unresponsive
\end{itemize}

\subsubsection{Implementation Testing Functionalities}
\begin{itemize}
    \item Change architecture of fall detection model
    \item Change amount of labelling lag (for fall prevention models)
\end{itemize}

\subsection{Usability}

\subsubsection{User Requirements}
\begin{itemize}
    \item Phone with a cellular connection (to contact medical assistance)
    \item Have their phone and chest strap on at all times
    \item Have sufficient battery on their phone and chest strap
\end{itemize}

\subsection{Design}
A dark mode style design with purple accents was selected for the app, given these are calming colours, and dark mode is far less straining on the eyes.

SwiftUI was used to develop the interface and create custom style classes to ensure consistent styling throughout the entire app.

\begin{figure}[ht]
\centering
\subfigure{%
\includegraphics[width=0.15\textwidth]{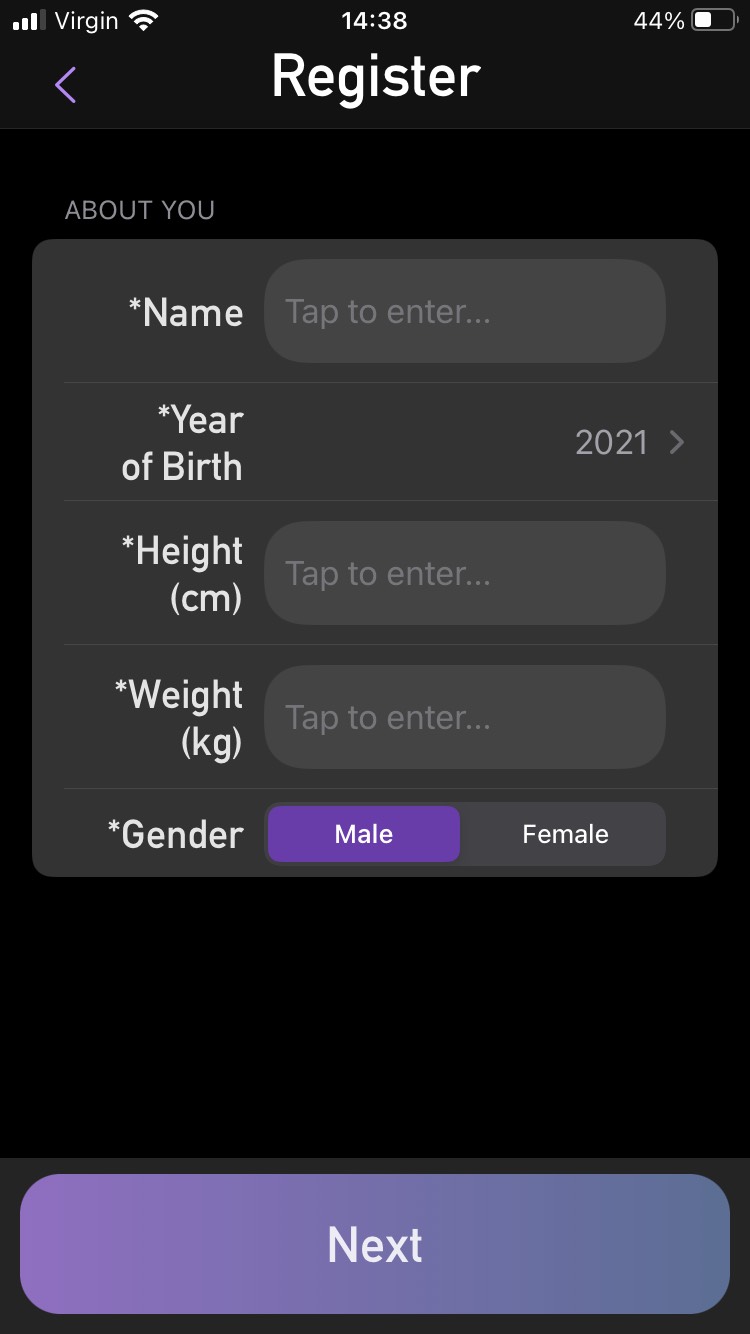}
\label{fig:fd-reg0}}
\quad
\subfigure{%
\includegraphics[width=0.15\textwidth]{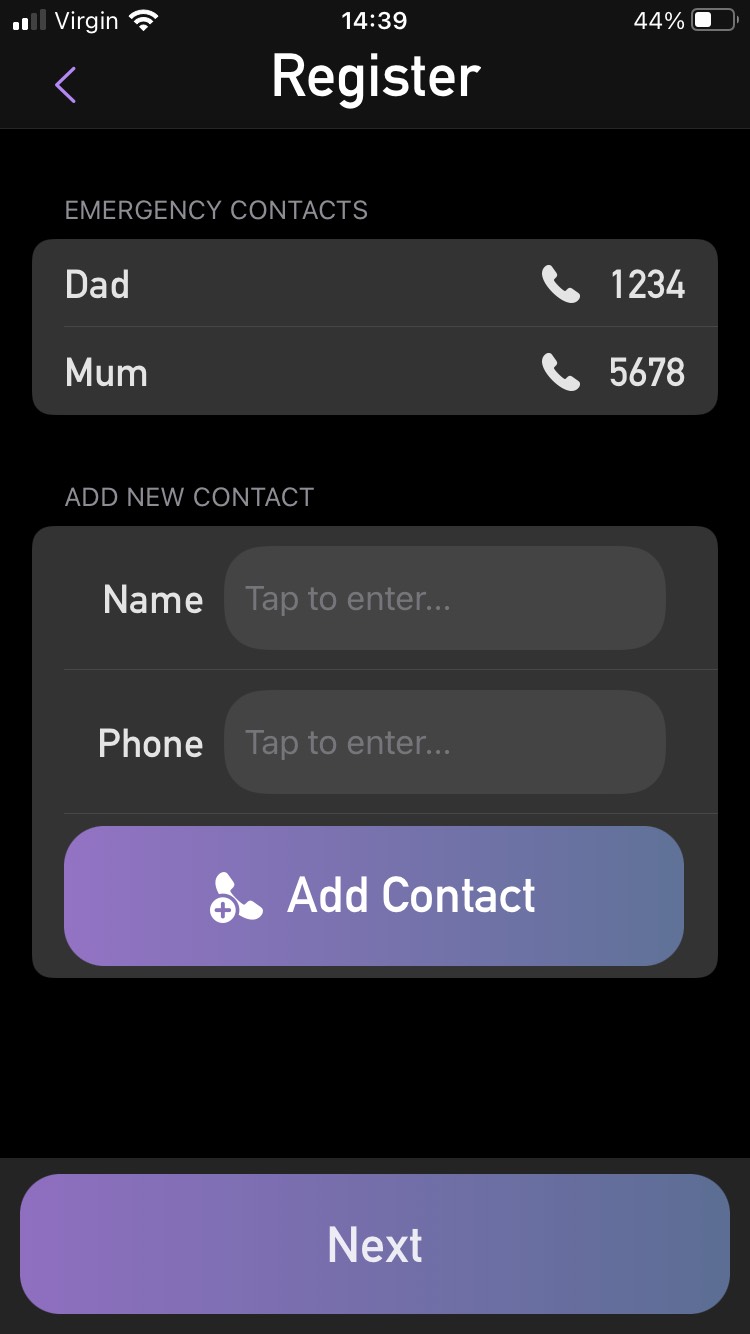}
\label{fig:fd-reg1}}
\quad
\subfigure{%
\includegraphics[width=0.15\textwidth]{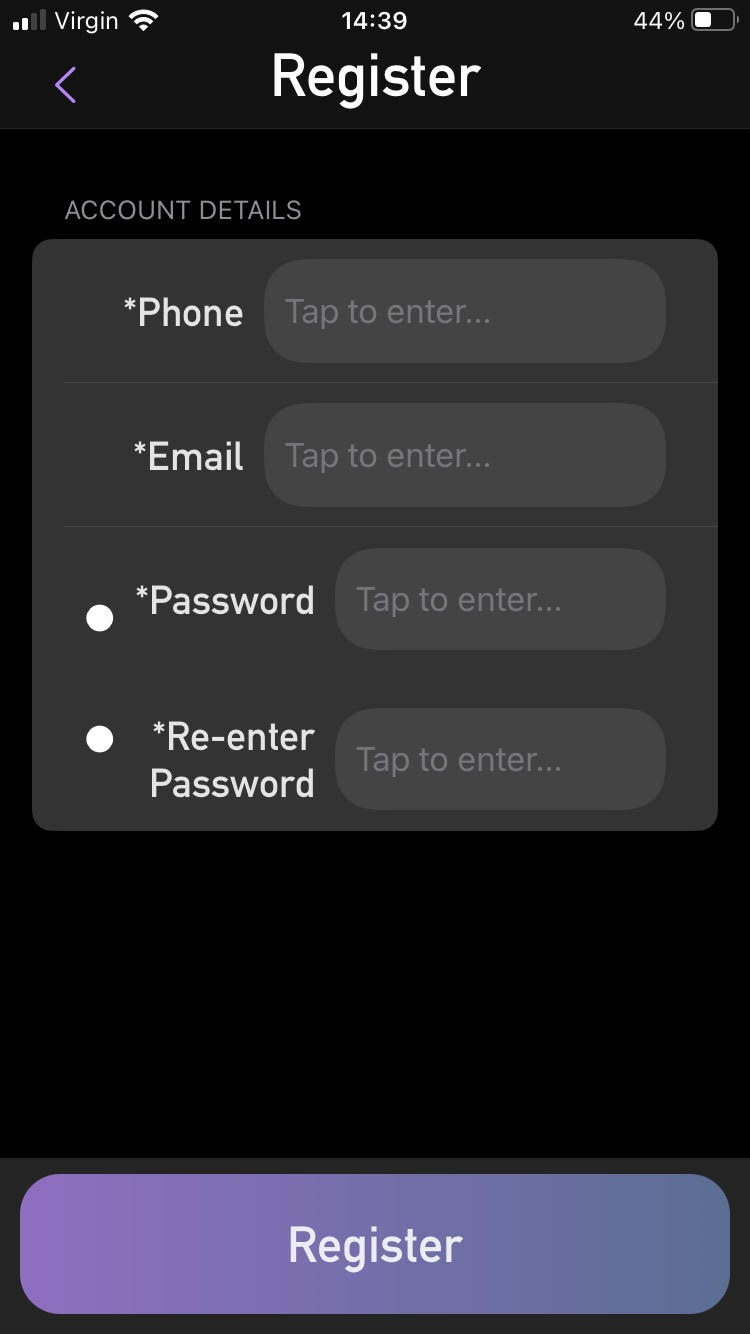}
\label{fig:fd-reg2}}
\caption{Registration pages}
\end{figure}

\begin{figure}[ht]
\centering
\subfigure{%
\includegraphics[width=0.15\textwidth]{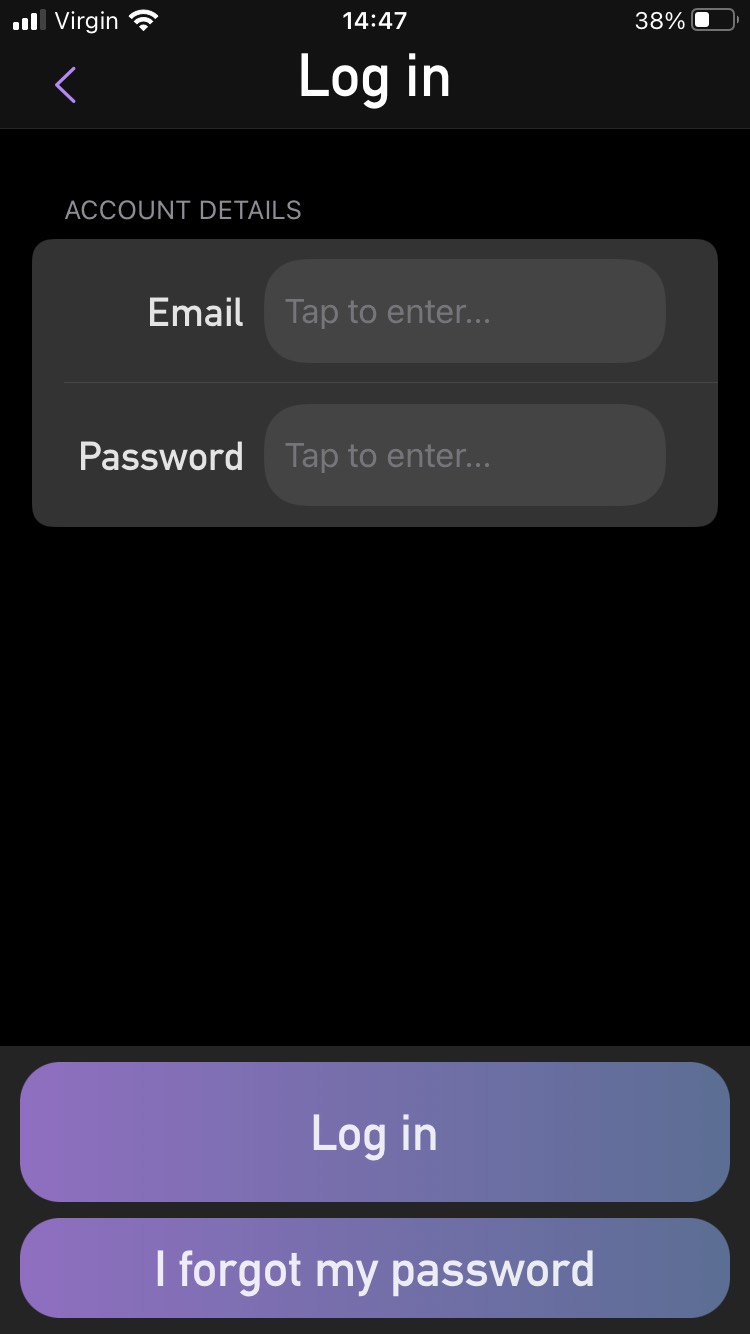}
\label{fig:fd-login}}
\quad
\subfigure{%
\includegraphics[width=0.15\textwidth]{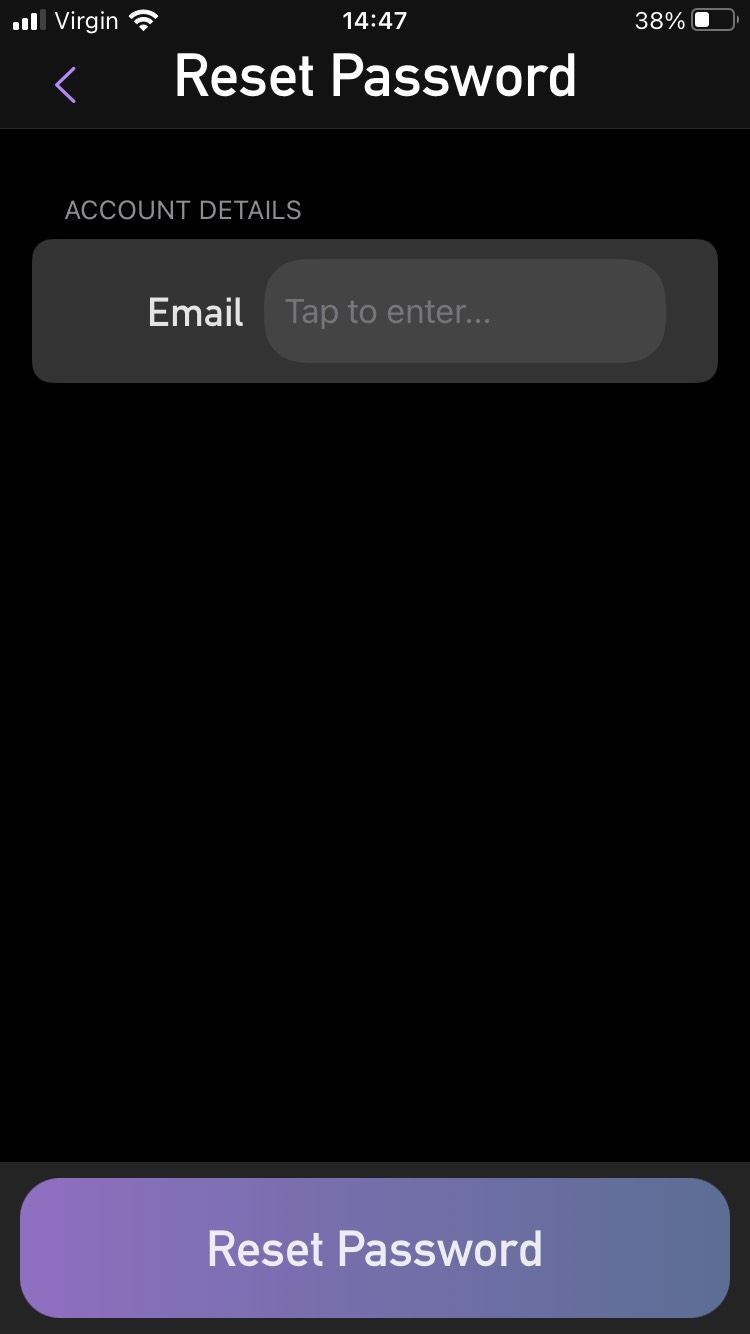}
\label{fig:fd-resetpass}}
\quad
\subfigure{%
\includegraphics[width=0.15\textwidth]{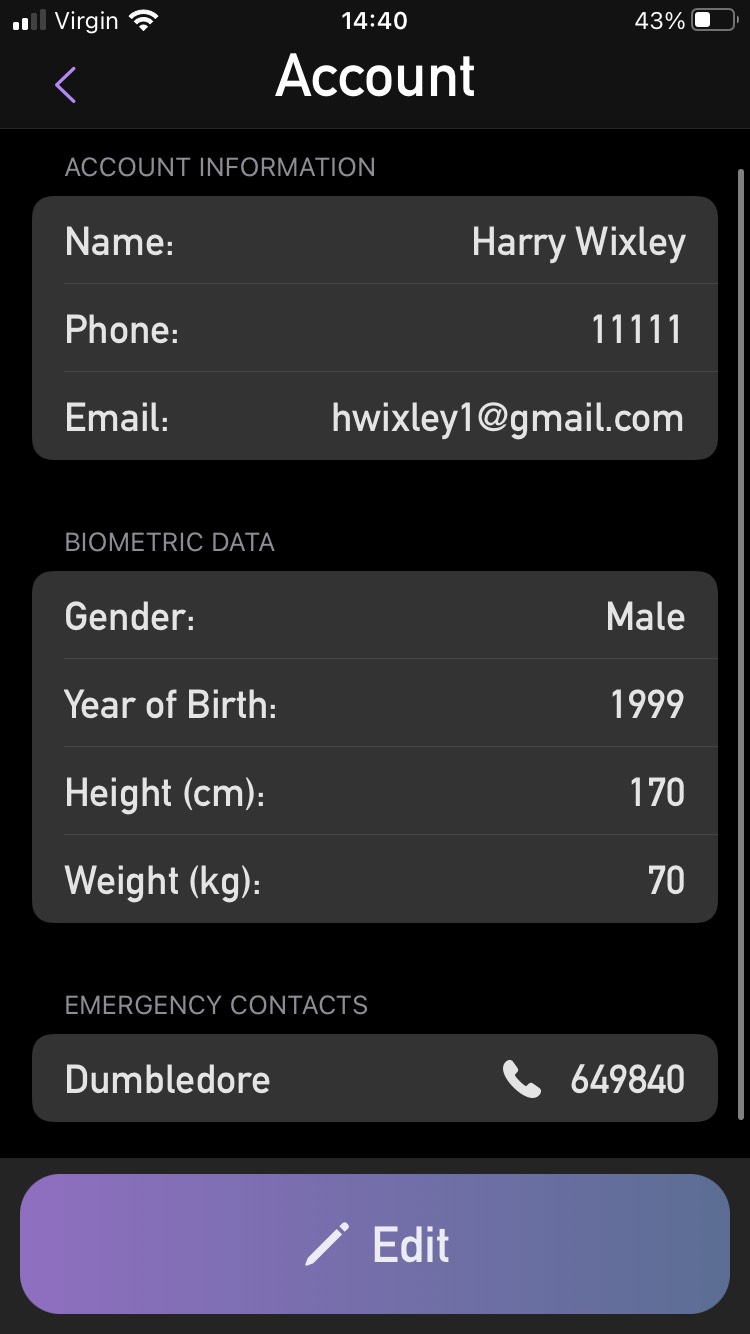}
\label{fig:fd-account}}
\subfigure{%
\includegraphics[width=0.15\textwidth]{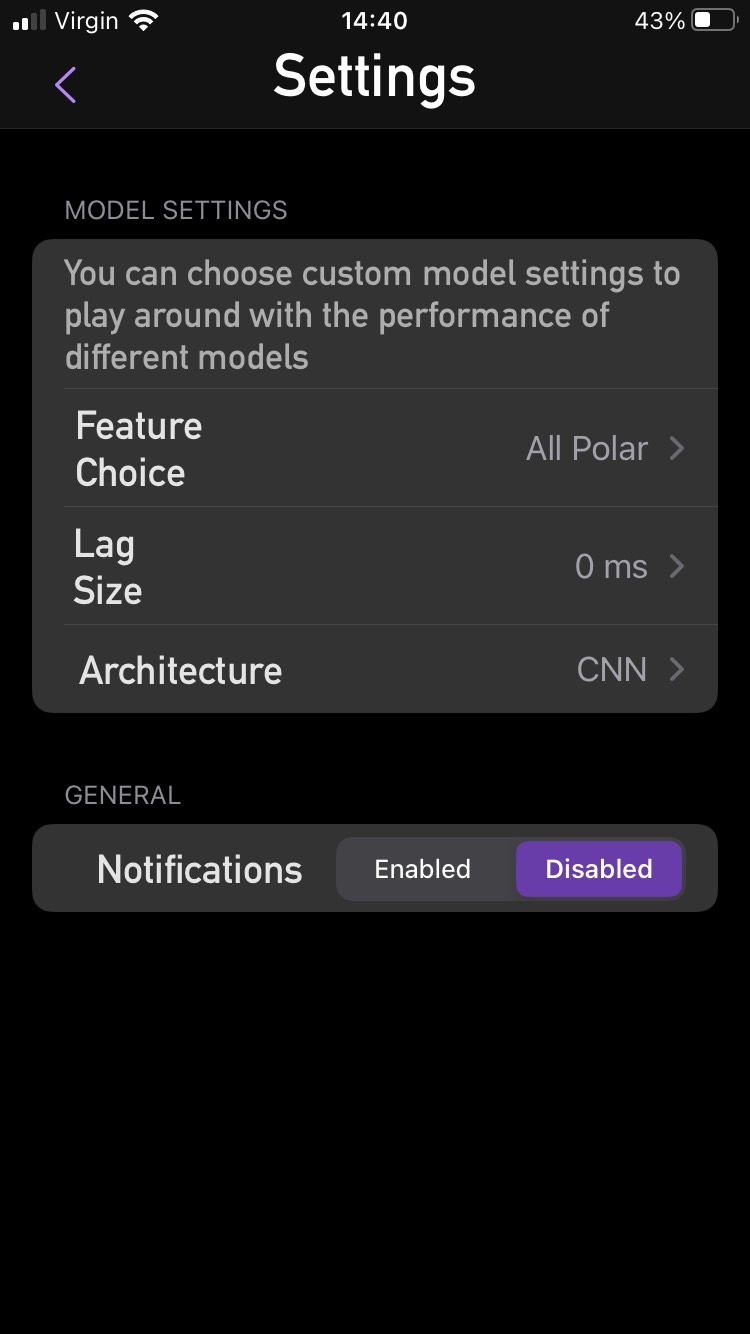}
\label{fig:fd-settings}}
\caption{Login, reset password, account, and settings pages}
\end{figure}

\begin{figure}[ht]
\centering
\subfigure{%
\includegraphics[width=0.15\textwidth]{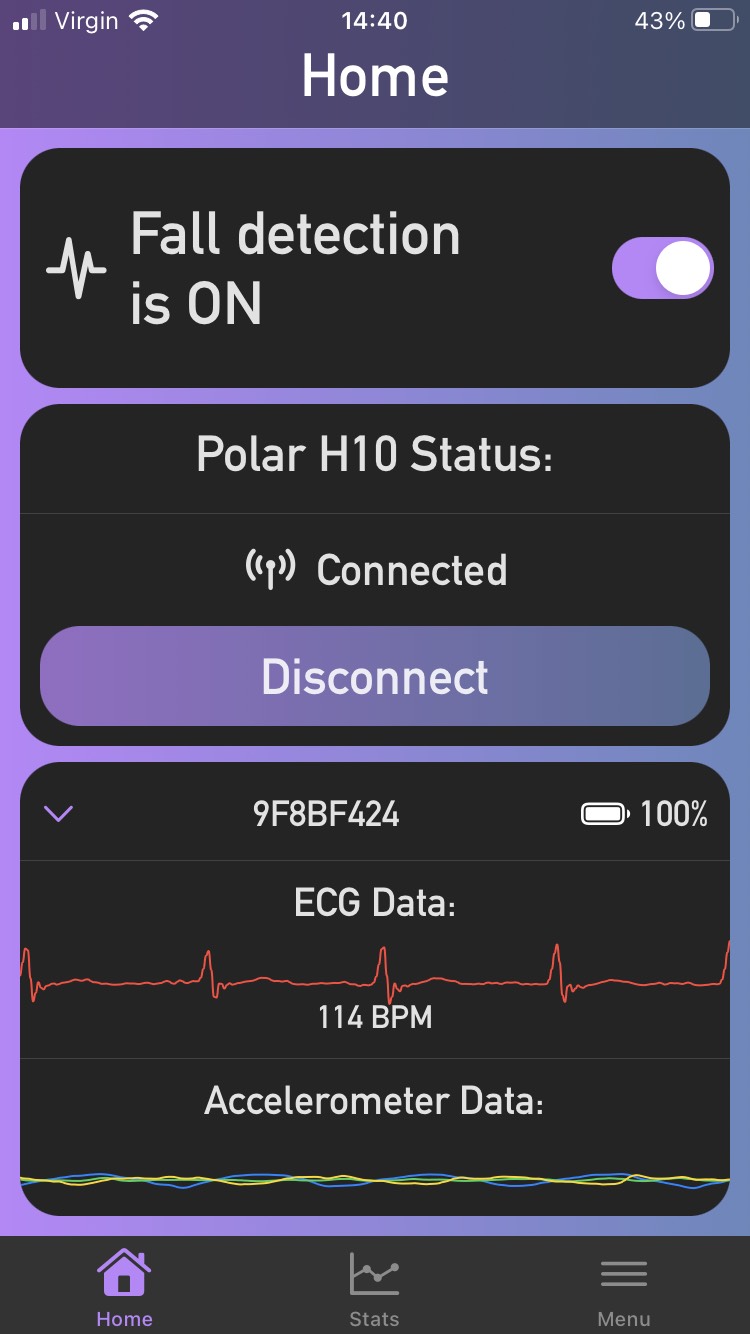}
\label{fig:fd-disconnected}}
\quad
\subfigure{%
\includegraphics[width=0.15\textwidth]{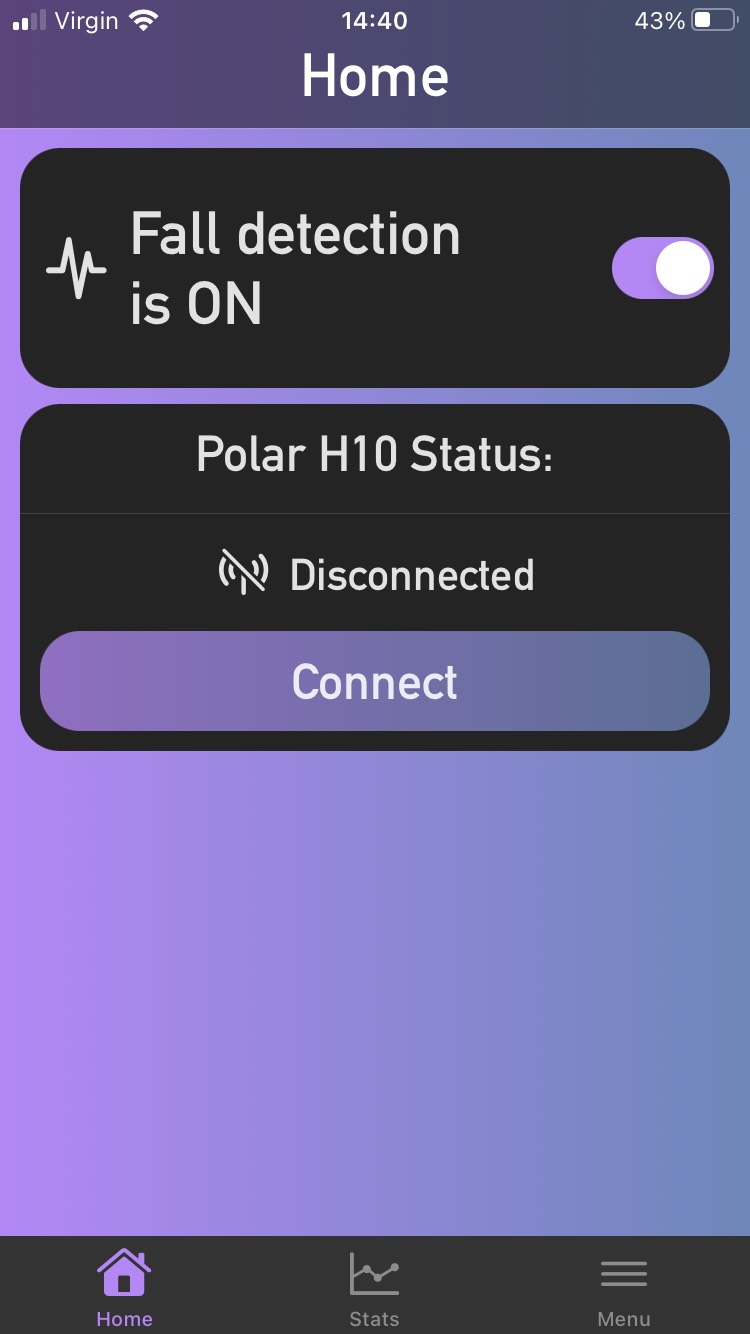}
\label{fig:fd-connected}}
\quad
\subfigure{%
\includegraphics[width=0.15\textwidth]{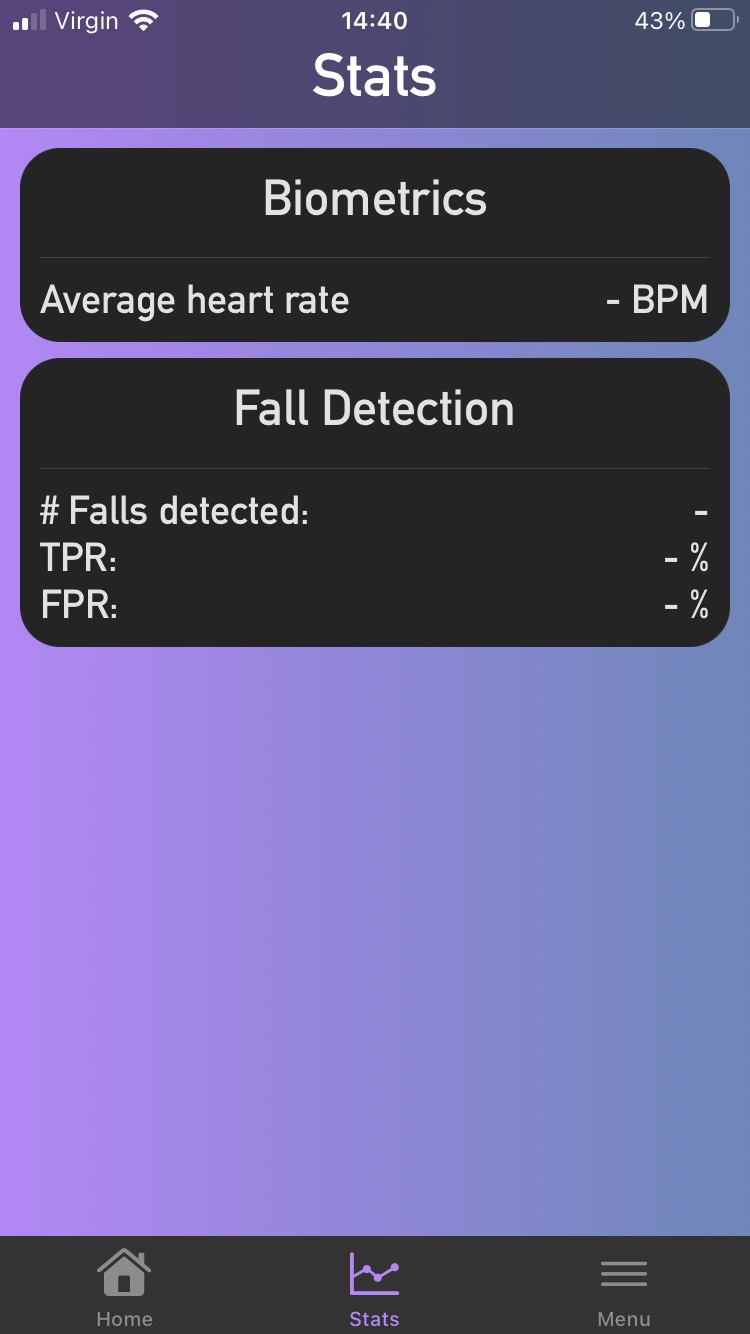}
\label{fig:fd-connected}}
\caption{Main app page when the Polar H10 is connected/disconnected, and the stats page. Please note the stats page had not been connected to the backend and thus statistics were not saved in the cloud (explaining the empty statistic values).}
\end{figure}

\section{Security \& Privacy}

\subsection{Database}
Database rules were implemented meaning authenticated users (those who are logged in) can only read/write their own data, and have no access to other users' data.


\subsection{App Privileges Requirements}
\begin{itemize}
    \item \underline{Always on location:} required to send a user's location to their emergency contacts if a fall is detected and the user is unresponsive
    \item \underline{Background processing:} required to run live fall detection
    \item \underline{Always on bluetooth access:} required to connect to the Polar H10 device
\end{itemize}

\section{Real Time Machine Learning}



\subsection{Model Customisation}
In the Settings section of the app I have added buttons allowing users to customise the fall detection model to be run, by choosing the: model lag size (for varying between detection and prevention models) and model architecture. This was developed purely as a feature for implementation testing.

\subsection{Live Preprocessing}
Any preprocessing done to the training data of our fall detection model must also be performed on the live sensor data to ensure it is represented in a way that the model understands.

This can evidently add a lot of computational complexity to the execution of the model. Thus it is very important to balance the computational complexity with model complexity. This is particularly difficult when combining the domain of safety-critical systems with live data processing.

In this case, due to the data not being unnecessarily large there was little latency with performing the preprocessing, however, it did make the app energy hungry.

\subsection{Live Model Inference}
This was performed using the TensorFlowLite Swift library. This was executed using the following steps: retrieve the live data for a specific window size, import the tflite models into my app, import the preprocessing transformation parameters into my app (ie. mean and std. dev. of training features for data standardisation), and perform model inference once the live data has been collected and preprocessed.

To retrieve the live data a timer was used to retrieve the sensor data over 100ms intervals. Once this data has been collected from the latest interval this interval data is appended to a matrix with the last $w-1$ intervals (representing a sample of size $(w,75)$ given there are 75 features in a 100ms interval). This matrix is then standardised (using the imported mean and std. dev. parameters) and sent to the inference code for processing.

Once the data is sent to the inference code the loaded model's interpreter is invoked on the input data. The main difficulty with this step was converting the model's output Tensors into Swift's native Data datatype given there was no available software for dealing with such a conversion. Due to this pointers had to be used manually to retrieve the output Tensor bytes and parse it into a Data object.

\section{Background Processing}
In order to constantly be able to detect falls (even if the user's phone is locked) we need to be able to run background processes. Given background processes are energy and performance hungry (given they are run constantly) any code to be run in the background must be dispatched to concurrent thread queues for optimal efficiency (given phones utilise multicore processors). 

Thread process allocation and concurrency was initially difficult. However, after some research and testing successful execution was achieved by dispatching the live preprocessing and model inference code to the background thread process queue. 

\section{Emergency Notifications}
Given we are providing notifications for critical emergency situations we must ensure they are robust and useful. Text messages were chosen as a means to contact emergency contacts as this method is most likely to get a quick response given most people carry their phones with them everywhere, and text messages do not require an internet connection to be sent.

To ensure reliable notifications it was decided to integrate Apple Push Notifications. This allowed for local notifications (issued by the fall detection code run in background processes) to be sent to the user if a fall is detected with buttons to reply ("I am fine" and "Send help"). If no response is detected by the user after 1 minute the user's emergency contacts are notified.

\subsection{Text Messages}
A function was configured that sends text messages by connecting to the Twilio API using custom requests in Alamofire. As discussed above this functionality was used to notify a users' emergency contacts that the user has taken a fall. This text message includes the user's coordinates (longitude and latitude), and the closest postal address to ensure the emergency contact can reliably locate the user.

\subsection{Apple Push Notifications}
Local APNs were configured in order to notify a user if a fall is detected.

These added a significant amount of complexity to development due to the requirement of Apple certificates, and background notification handling.


\chapter{Discussion \& Concluding Remarks}

\section{Discussion}
The model deployed through the smart application produced high inference accuracy showing that the data collection, processing, and modelling techniques used in this study were valid.

The best performing model overall was the ResNet152 on a standardised, and shuffled dataset with window size 20 which achieved 92.8\% AUC, 87.28\% sensitivity, and 98.33\% specificity. It must be noted this model had 58,141,634 trainable parameters resulting in a large complexity. Although the inference time was still fast on the iPhone (after being converted into a tflite model) it resulted in the app having a high power consumption rating (inspected using the Xcode testing panel).

Due to the usage of such a complex ResNet architecture, and the success of smaller ResNets on other fall detection datasets I would have expected an even higher AUC score. However, I believe this can be potentially attributed to multiple factors that hindered the quality of the dataset, including: the quality of dataset labelling, the relatively small amount of dataset samples, the small test subject representation, and the lack of using frequency domain processing techniques. This is further supported by the poor generalisation of the baseline and LSTM results. Such poor generalisation seems to indicate that there must be issues with the data. Given only complex models can generalise well this is potentially due to the fact these models have learned to handle the aforementioned dataset limitations/inaccuracies (such as learning the labelling latency).

Considering these dataset limitations it is likely in future dataset iterations that higher performance could be achieved, and lower complexity models could be used.



Besides fixing the dataset limitations, one of the main ways this project could be expanded further would be to investigate the development of more refined and customised hardware for the fall detection processing/inference. Doing this processing on a phone was very energy hungry and thus not realistic for a commmercial system. For example, a technique such as reservoir computing\footnote{A machine learning algorithm for processing information generated by dynamical systems using observed time-series data (\cite{gauthier_bollt_griffith_barbosa_2021}).} could pose benefits in computational efficiency if optimised on the deployed fall detection algorithm.



\section{Conclusion}
In conclusion, this study developed multiple application components in order to investigate the development challenges and choices for fall detection systems. The smart application proved to be successful based on results from fall detection modelling experiments and model mobile deployment.

The best performing model overall was the ResNet152 on a standardised, and shuffled dataset with window size 20 which achieved 92.8\% AUC, 87.28\% sensitivity, and 98.33\% specificity. Given these results it is evident that accelerometer and ECG sensors combined with personalised data are beneficial for fall detection, and allow for the discrimination between complex ADLs (ie. tying shoelaces) and falls.

For this ResNet152 architecture with these sensor inputs it was determined that the most useful directions for further research would be adapting the dataset labelling protocol, increasing the number of dataset samples and subjects, and using frequency domain processing. By improving the quality of the data we could increase the model's ability to generalise, thus implying such research directions could produce less complex models with similar/better AUC scores.

\section{Recommendations for Future Work}

\subsection{Alternative Labelling Method - Critical Phase}
This refers to labelling that only marks the critical phase of a fall. This would allow us to use smaller window sizes implying less input features and ultimately lower complexity.

\subsection{Include Results of Deployed Model Inference Times}
This would be extremely useful to gauge and compare the complexity of varying models as a means to evaluate their reliability and suitability for fall detection in a commercial system.

\subsection{Alternative ECG Sensor Setup}

\subsubsection{Develop a small standalone wearable device}
The main limitation of this project along with the majority of biomedical systems is that of the usability. Although it is impossible to remove wearable devices completely (however this of course not the case of vision-based systems) we can try make them as minimal as possible, so they do not require regular charging and are not imposing on the appearance/comfort of the user.

This could be done by developing a self-contained system using analog nanotechnology. Analog chips are particularly relevant in this context due to their superior energy usage over digital chips, and the fact we are working with waveform data (accelerometer, and ECG sensors).

We could put all the software on this chip and use techniques such as reservoir computing to optimise it for computing ML algorithms.

With regards to specific products a ring could be considered due to its tiny footprint, and the fact it can easily be customised/designed to blend into (or even stand out from) a user's outfit.

\subsubsection{Use implant ECG sensors}
This could be useful for elderly with existing ECG implant sensors, as this would not require them to wear any additional external devices.

\subsubsection{Use optical ECG sensors on a smart watch}
This could be useful to prevent the requirement for the user to wear a chest strap constantly and have their phone on them. Rather it could all be integrated on a smart watch (ie. Apple Watch).

\subsection{Use other biometric readings}
\begin{itemize}
    \item Blood pressure - this could be extremely useful to detect whether users were about to faint by a significant drop/elevation in blood pressure.
    \item Blood glucose - this would be particularly useful for users with diabetes as it could detect whether users were about to faint due to a significant change of blood glucose. It could also include functionality to alert the user if their blood glucose drops below a certain threshold.
    \item Blood oxygen
    \item Respiration rate
    \item Skin temperature
    \item Extra user features: advanced biometrics (fat mass, bone mass, muscle mass, water), and medical conditions. 
\end{itemize}

\subsection{Abnormal Gait Detection}
Use this data to develop an abnormal gait detection model as a means to detect onset neurological conditions. For example detecting the freezing of gait experienced by a Parkinson's patient.

\printbibliography[heading=bibliography]

\appendix

\chapter{Technical Materials}

\section{Source Code}
All the source code for this project can be found \textbf{\href{https://github.com/hwixley-honours-project}{here}}, which include the following repositories:
\begin{itemize}
    \item \textbf{\href{https://github.com/hwixley-honours-project/firebase-fddg}{firebase-fddg (Swift)}}: an iOS data collection app that stores all data on Firebase's Firestore. This app was not used due to data protection reasons.
        \begin{itemize}
            \item \textbf{Swift Dependencies:} Polar SDK (for connecting to Polar devices), Firebase SDK (for storing the data in a NoSQL database on Firebase's cloud service)
        \end{itemize}
    
    \item \textbf{\href{https://github.com/hwixley-honours-project/localhost-fddg}{localhost-fddg (Swift)}}: an iOS data collection app that connects to a custom LAN server, and sends data to the MongoDB database connected to this server.
        \begin{itemize}
            \item \textbf{Swift Dependencies:} Polar SDK (for connecting to Polar devices), Alamofire SDK (for performing networking in Swift such as API calls or custom server requests)
        \end{itemize}
    
    \item \textbf{\href{https://github.com/hwixley-honours-project/localhost-data-collection}{localhost-data-collection (JavaScript)}}: a secure NodeJS server that connects to a MongoDB database for data storage.
        \begin{itemize}
            \item \textbf{NodeJS Dependencies:} Express (for creating the server), Mongoose (for connecting to the MongoDB instance), BodyParser (for setting up the JSON parser and increasing the maximum request body size to 50MB), Express IP Access Control (for creating an express middleware for IP-based access control)
        \end{itemize}
    
    \item \textbf{\href{https://github.com/hwixley-honours-project/localhost-data-preprocessing}{localhost-data-preprocessing (Python)}}: includes Jupyter notebooks for all preprocessing techniques (JSON parsing, feature scaling, lag features, sliding windows, and denoising autoencoders), a notebook for testing this data on library ML models (ie. scikit-learn models) for preprocessing evaluation, notebooks for deep learning implementations using PyTorch (ie. ResNet, CNN, LSTM), and a notebook for exporting PyTorch models to .tflite for mobile phone integration.
        \begin{itemize}
            \item \textbf{Python Dependencies:} Numpy (for numerical processing), BSON (for BSON file parsing), Matplotlib (for data visualizations), PyTorch (for developing the deep learning models), ONNX \& TensorFlow (for PyTorch model exporting to tflite)
        \end{itemize}
    
    \item \textbf{\href{https://github.com/hwixley-honours-project/fall-detector-app}{fall-detector-app (Swift)}}: an iOS app for fall detection. This app takes the learnt fall detection model and uses it against live sensor inputs for fast and reliable fall detection. I have also added login and create account functionality using Firebase to allow users to store data such as emergency contacts and more.
        \begin{itemize}
            \item \textbf{Swift Dependencies:} Polar SDK (for connecting to Polar devices), Firebase SDK (for connecting to a NoSQL database in the cloud, and performing user authentication), TensorFlowLite (for running tflite models on my app), Alamofire (for performing API requests to Twilio in order to send emergency text messages)
        \end{itemize}
\end{itemize}

\section{Dataset}
Please contact my supervisor, Kianoush Nazarpour (kianoush.nazarpour@ed.ac.uk), if you would like access to the dataset(s).


\chapter{Ethics Materials}



\section*{Supplementary material}

## B.1 Data Protection for Research

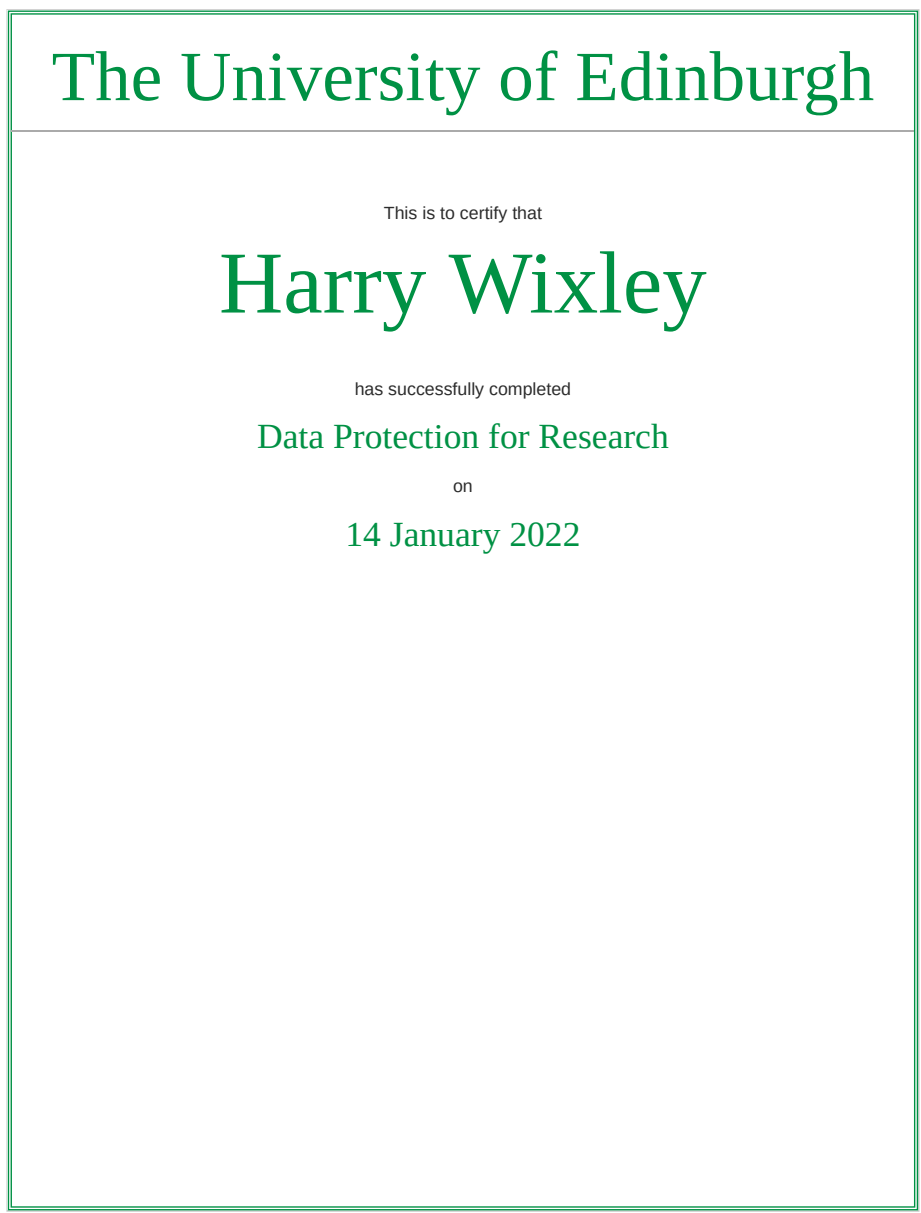

## B.2 Participants' information sheet

**Participant Information Sheet**

| Project title: | Smart Wearable Sensors for Fall Detection |
|---|---|
| Principal investigator: | Kianoush Nazarpour |
| Researcher collecting data: | Harry Wixley |

This study was certified according to the Informatics Research Ethics Process, RT number 18048. Please take time to read the following information carefully. You should keep this page for your records.

**Who are the researchers?**

-Harry Wixley

-Kianoush Nazarpour

**What is the purpose of the study?**

The aim of this study is to collect movement and heart rate data using a smartphone in order to develop a dataset for a machine learning based fall detection system.

**Why have I been asked to take part?**

The research target group is not tailored in any specific way.

**Do I have to take part?**

No – participation in this study is entirely up to you. You can withdraw from the study at any time, without giving a reason. Your rights will not be affected. If you wish to withdraw, contact the PI. We will stop using your data in any publications or presentations submitted after you have withdrawn consent. However, we will keep copies of your original consent, and of your withdrawal request.

**What will happen if I decide to take part?**

- <u>Kinds of data being collected:</u>
    - o Movement Data
        - Accelerometer data from the heart rate sensor
        - Accelerometer data, gyroscope data, and magnetometer data from a smartphone

THE UNIVERSITY *of* EDINBURGH
**informatics**

- Biometric Data
    - Heart rate: ECG (electrocardiogram) signal data retrieved from the Polar H10 heart rate sensor
    - Height
    - Weight
    - Age
    - Gender
    - Medical conditions
- <u>Means of collection:</u>
    - Biometric data (excluding ECG data) will be collected in-app
    - Movement and ECG data will be collected in a large space with soft matting (such as a gymnasium) to allow participants to fall safely (protective equipment can also be provided such as a helmet, knee pads, and elbow pads)
- <u>Duration of session:</u>
  30 minutes per participant
- <u>Is participant audio/video is being recorded:</u>
  Yes some video will be recorded as a means to display how this data was collected
- <u>How often, where, when:</u>
  There will only need to be one session.

**Are there any risks associated with taking part?**

Yes there is a risk of physical injury by participating in this study. This study requires participants to purposefully fall over so we can collect data of what a fall looks like. However, many safety measures will be put into place to ensure the safety of participants. This study will be conducted in a space with soft matting to ensure these falls do not hurt participants, on top of this protective gear will be provided (such as a helmet, knee pads, and elbow pads).

**Are there any benefits associated with taking part?**

A chocolate bar will be provided to all participants at the end of the session.

THE UNIVERSITY of EDINBURGH
informatics

**What will happen to the results of this study?**
The results of this study may be summarised in published articles, reports and presentations. Quotes or key findings will be anonymized: We will remove any information that could, in our assessment, allow anyone to identify you. With your consent, information can also be used for future research. Your data may be archived for a minimum of 2 years.

**Data protection and confidentiality.**
Your data will be processed in accordance with Data Protection Law. All information collected about you will be kept strictly confidential. Your data will be referred to by a unique participant number rather than by name. Your data will only be viewed by the researcher/research team (Harry Wixley and Kianoush Nazarpour).

All raw electronic data will be stored on a password-protected encrypted computer, on the School of Informatics' secure file servers, or on the University's secure encrypted cloud storage services (DataShare, ownCloud, or Sharepoint) and all paper records will be stored in a locked filing cabinet in the PI's office. In future this raw electronic data will be used to train a machine learning model which will be stored on Google's Firebase server (allowing us to execute model testing in the cloud to minimise execution time and power consumption). This implies that although all the learned features extracted from the dataset in training time will be stored on this service it does not pose any risk to your data as it would be infeasible to extrapolate it from the model's learned features. Furthermore, model updating will be performed via a technique called federated learning implying the model will be updated locally using the local data samples to ensure that this data cannot be accessed online. Your consent information will be kept separately from your responses in order to minimise risk.

**What are my data protection rights?**
The University of Edinburgh is a Data Controller for the information you provide. You have the right to access information held about you. Your right of access can be exercised in accordance Data Protection Law. You also have other rights including rights of correction, erasure and objection. For more details, including the right to lodge a complaint with the Information Commissioner's Office, please visit

www.ico.org.uk. Questions, comments and requests about your personal data can also be sent to the University Data Protection Officer at dpo@ed.ac.uk.
For general information about how we use your data, go to: edin.ac/privacy-research

**Who can I contact?**
If you have any further questions about the study, please contact the lead researcher, [Harry Wixley, s1803764@ed.ac.uk].
If you wish to make a complaint about the study, please contact inf-ethics@inf.ed.ac.uk. When you contact us, please provide the study title and detail the nature of your complaint.

**Updated information.**
If the research project changes in any way, an updated Participant Information Sheet will be made available on http://web.inf.ed.ac.uk/infweb/research/study-updates.
[**NB:** the PI should notify the Ethics panel on inf-ethics@ed.ac.uk to upload any update PIS to the website]

**Consent**
By proceeding with the study, I agree to all of the following statements:
- I have read and understood the above information.
- I understand that my participation is voluntary, and I can withdraw at any time.
- I consent to my anonymised data being used in academic publications and presentations.
- I allow my data to be used in future ethically approved research.
    - ☐ I agree

## B.3 Participants' consent form

Participant number: _____

**Participant Consent Form**

| | |
|---|---|
| Project title: | Smart Wearable Sensors for Fall Detection |
| Principal investigator (PI): | Kianoush Nazarpour |
| Researcher: | Harry Wixley |
| PI contact details: | kianoush.nazarpour@ed.ac.uk |

By participating in the study you agree that:

- I have read and understood the Participant Information Sheet for the above study, that I have had the opportunity to ask questions, and that any questions I had were answered to my satisfaction.

- My participation is voluntary, and that I can withdraw at any time without giving a reason. Withdrawing will not affect any of my rights.

- I consent to having my movement and biometric data (as described in the participant information sheet) being recording.

- I consent to my anonymised data being used in academic publications and presentations.

- I understand that my anonymised data will be stored for the duration outlined in the Participant Information Sheet.

**Please tick yes or no for each of these statements.**

1. I agree to having video recorded.  
   ☐ Yes ☐ No

2. I allow my data to be used in future ethically approved research.  
   ☐ Yes ☐ No

3. I agree to take part in this study.  
   ☐ Yes ☐ No

Name of person giving consent  Date dd/mm/yy  Signature
_____  _____  _____

Name of person taking consent  Date dd/mm/yy  Signature
_____  _____  _____

THE UNIVERSITY of EDINBURGH
informatics



\end{document}